%% file: MAIN.tex
\documentclass[11pt]{article}
\usepackage[a4paper,hmargin=1.0in,vmargin=1.0in]{geometry}
\usepackage{mathtools,amsthm,amssymb,setspace,sectsty,bbm}%,lmodern}
\usepackage[round]{natbib}
\usepackage[usenames,dvipsnames]{xcolor}
\usepackage[linktocpage=true,pagebackref=true,colorlinks,allcolors=blue,bookmarks=true,bookmarksopen,bookmarksnumbered]{hyperref}

\newtheoremstyle{DStheorem}% name of the style to be used
  {\topsep}% measure of space to leave above the theorem. E.g.: 3pt
  {\topsep}% measure of space to leave below the theorem. E.g.: 3pt
  {\itshape}% name of font to use in the body of the theorem
  {0pt}% measure of space to indent
  {\scshape}% name of head font
  {.}% punctuation between head and body
  { }% space after theorem head; " " = normal interword space
  {\thmname{#1}\thmnumber{ #2}\thmnote{ (#3)}}
\theoremstyle{DStheorem}
\newtheorem{theorem}{Theorem}[section]
\newtheorem{lemma}[theorem]{Lemma}
\newtheorem{claim}[theorem]{Claim}
\newtheorem{observation}[theorem]{Observation}

\let\oldproofname=\proofname
\renewcommand{\proofname}{\rm\sc{\oldproofname}}

\newcommand{\bstitle}[1]{\texorpdfstring{$\boldsymbol{#1}$}{}}
\newcommand{\MyAbove}[2]{\genfrac{}{}{0pt}{}{#1}{#2}}
\newcommand{\eoq}{\mathrm{EOQ}}
\newcommand{\potwo}{\mathrm{TB01}}
\newcommand{\mylcm}{\mathrm{LCM}}
\newcommand{\myLB}{\mathrm{LB}}
\newcommand{\bs}[1]{\boldsymbol{#1}}
\newcommand{\bbR}{\mathbbm{R}}
\newcommand{\eps}{\epsilon}
\newcommand{\poly}{\mathrm{poly}}
\newcommand{\opt}{\mathrm{OPT}}
\newcommand{\ex}[1]{\mathbbm{E}\left[#1\right]}
\newcommand{\expar}[1]{\mathbbm{E}[#1]}
\newcommand{\exsub}[2]{\mathbbm{E}_{#1}\left[#2\right]}
\newcommand{\exsubpar}[2]{\mathbbm{E}_{#1}[#2]}
\newcommand{\pr}[1]{\mathrm{Pr}\left[#1\right]}
\newcommand{\prpar}[1]{\mathrm{Pr}[#1]}

\sectionfont{\large} \subsectionfont{\normalsize}
\allowdisplaybreaks
\makeindex

\begin{document}

\begin{titlepage}

\title{Improved Approximation Guarantees for \\
Joint Replenishment in Continuous Time}
\author{%
Danny Segev\thanks{School of Mathematical Sciences and Coller School of Management, Tel Aviv University, Tel Aviv 69978, Israel. Email: {\tt segevdanny@tauex.tau.ac.il}. Supported by Israel Science Foundation grant 1407/20.}}
\date{}
\maketitle

\setcounter{page}{200}
\thispagestyle{empty}

\begin{abstract}
The primary objective of this work is to revisit and revitalize one of the most fundamental models in deterministic inventory management, the continuous-time joint replenishment problem. Our main contribution consists of resolving several long-standing open questions in this context. For most of these questions, we obtain the first quantitative improvement over power-of-$2$ policies and their nearby derivatives, which have been state-of-the-art in terms of provable performance guarantees since the mid-80's.
\end{abstract}


\end{titlepage}

% CONTENTS %%%%%%%%%%%%%%%%%%%%%%%%%%%%%%%%%%%%%%%
\setcounter{page}{200}
\pagestyle{empty}
\tableofcontents

% SECTIONS %%%%%%%%%%%%%%%%%%%%%%%%%%%%%%%%%%%%%%%%%%%%%%%%%%%%%%%%%%%
\newpage
\pagestyle{plain}
\setcounter{page}{1}
\input{TEX-Intro.tex}
\input{TEX-EPTAS-Variable-Base}
\input{TEX-Black-Box-Reduction}
\input{TEX-Evenly-Spaced}
\input{TEX-Resource-Constrained}
\input{TEX-Remarks.tex}

% BIB %%%%%%%%%%%%%%%%%%%%%%%%%%%%%%%%%%%%%%%%%%%
\addcontentsline{toc}{section}{Bibliography}
\bibliographystyle{plainnat}
\bibliography{MAIN}

% APPENDIX %%%%%%%%%%%%%%%%%%%%%%%%%%%%%%%%%%%%%%
\appendix
\input{TEX-More-Proofs}

\end{document}

%% file: TEX-Intro.tex
\section{Introduction} \label{sec:intro}

The primary objective of this paper is to revisit and revitalize one of the most fundamental models in deterministic inventory management, the continuous-time joint replenishment problem. As it turns out, our recent work along these lines \citep{Segev23JRP} is by no means the final chapter in decades-long investigations of this classical paradigm. Rather, the current paper will present improved algorithmic ideas for cornerstone models in this context, along with new ways of analyzing their performance guarantees. Additionally, some of our findings serve as proofs of concepts, intended to open the door for fine-grained improvements in future research.

As second-year undergrad students quickly find out, and as supply-chain experts repeatedly rediscover, the joint replenishment problem has been an extremely versatile petri dish in developing the theoretical foundations of inventory management. Of no lesser importance is its role in boosting the practical appeal of this academic field across a massive body of work dating back to the mid-60's, with indirect explorations surfacing even earlier. For an in-depth appreciation of these statements, avid readers are referred to selected book chapters dedicated to this topic \citep{SilverP85, SimchiLeviCB, Zipkin00, MuckstadtS10}. At a high level, even though joint replenishment settings come in a variety of  flavors, they are all inherently concerned with the lot-sizing of multiple commodities over a given planning horizon, where our objective is to minimize long-run average operating costs. However, such settings routinely lead to challenging algorithmic questions about how numerous Economic Order Quantity (EOQ) models can be efficiently synchronized, as well as to long-standing analytical questions regarding their structural characterization. In a nutshell, on top of marginal  ordering and inventory holding costs, what makes such synchronization particularly problematic is the interaction between different commodities through joint ordering costs, incurred whenever an order is placed, regardless of its contents. To delve into the finer details of these questions and to set the stage for presenting our main contributions, we proceed by providing a formal mathematical description of the joint replenishment problem in its broadest continuous-time form. 

\subsection{Model formulation} \label{subsec:model_definition}

\paragraph{The Economic Order Quantity model.} For a gradual presentation, let us begin by introducing the Economic Order Quantity (EOQ) model, which will imminently form the basic building block of joint replenishment settings. Here, we would like to identify the optimal time interval $T$ between successive orders of a single commodity, aiming to minimize its long-run average cost over the continuous planning horizon $[0,\infty)$. Specifically, this commodity is assumed to be characterized by a stationary demand rate of $d$, to be completely met upon occurrence; in other words, we do not allow for lost sales or back orders. In this setting, periodic policies are simply those where our ordering frequency is uniform across the planning horizon. Namely, orders will be placed at time points $0, T, 2T, 3T, \ldots$, noting that the time interval $T$ is the sole decision variable to be optimized. To understand the fundamental tradeoff, on the one hand, each of these orders incurs a fixed cost of $K$ regardless of its quantity, meaning that we are motivated to place very infrequent orders. On the other hand, what pulls in the opposite direction is a linear holding cost of $h$, incurred per time unit for each inventory unit in stock, implying that we  wish to avoid high inventory levels via frequent orders. As such, the fundamental question is how to determine the time interval $T$, with the objective of minimizing long-run average ordering and holding costs. 

Based on the preceding paragraph, it is not difficult to verify that optimal policies will be placing orders only when their on-hand inventory level is completely exhausted, namely, zero inventory ordering (ZIO) policies are optimal. Therefore, we can succinctly represent the objective function of interest, 
\[ C(T) ~~=~~ \frac{ K }{ T } + HT \ , \]
with the convention that $H = \frac{ hd }{ 2 }$. The next claim gives a synopsis of several well-known properties exhibited by this function. We mention in passing that items~1-3 follow from elementary calculus arguments and can be found in most relevant textbooks; see, e.g.,  \citep[Sec.~7.1]{SimchiLeviCB} \citep[Sec.~3]{Zipkin00} \citep[Sec.~2]{MuckstadtS10}. In contrast, item~4 hides in various forms within previous literature, and we provide its complete proof in Appendix~\ref{app:proof_clm_EOQ_properties_4}.

\begin{claim} \label{clm:EOQ_properties}
The cost function $C : (0,\infty) \to \bbR_+$ satisfies the following properties:
\begin{enumerate}
    \item $C$ is strictly convex.

    \item The unique minimizer of $C$ is $T^* = \sqrt{ K / H }$.

    \item $C( \theta T^* ) = \frac{ 1 }{ 2 } \cdot (\theta + \frac{ 1 }{ \theta } ) \cdot C( T^* )$, for every $\theta > 0$.

    \item $\min \{ C( \alpha ), C( \beta  ) \} \leq \frac{ 1 }{ 2 } \cdot ( \sqrt{ \frac{ \beta }{ \alpha } } + \sqrt{ \frac{ \alpha }{ \beta } } ) \cdot C(T)$, for every $\alpha \leq \beta$ and $T \in [\alpha, \beta]$.
\end{enumerate}
\end{claim}

\paragraph{The joint replenishment problem.} With these foundations in place, the essence of joint replenishment can be captured by posing the next question: How should we coordinate multiple Economic Order Quantity models, when different commodities are tied together via joint ordering costs? Specifically, we wish to synchronize the lot sizing of $n$ distinct commodities, where each commodity $i \in [n]$ is coupled with its own EOQ model, parameterized by ordering and holding costs $K_i$ and $H_i$, respectively. As explained above, setting a time interval of $T_i$ between successive
orders of this commodity would lead to a marginal operating cost of $C_i(T_i) = \frac{ K_i }{ T_i } + H_i T_i$. However, the  caveat is that we are concurrently facing joint ordering costs, paying $K_0$ whenever an order is placed, regardless of its particular subset of commodities.

Given these ingredients, it is convenient to represent any joint replenishment policy through a vector $T = (T_1, \ldots, T_n)$, where $T_i$ stands for the time interval between successive
orders of each commodity $i \in [n]$. For such policies, the first component of our objective function encapsulates the sum of marginal EOQ-based costs, $\sum_{i \in [n]} C_i( T _i )$. The second component, which will be designated by $J(T)$, captures long-run average joint ordering costs. Formally, this term is defined as the asymptotic density
\begin{equation} \label{eqn:joint_cost_def}
J( T ) ~~=~~ K_0 \cdot \lim_{\Delta \to \infty} \frac{ N(T,\Delta) }{ \Delta } \ ,
\end{equation}
where $N(T,\Delta)$ stands for the number of joint orders occurring across $[0,\Delta]$ with respect to the time intervals $T_1, \ldots, T_n$.  That is, letting ${\cal M}_{T_i,\Delta} = \{ 0, T_i, 2T_i, \ldots, \lfloor \frac{ \Delta }{ T_i } \rfloor \cdot T_i \}$ be the integer multiples of $T_i$ within $[0,\Delta]$, we  have $N(T,\Delta) = | \bigcup_{i \in [n]} {\cal M}_{T_i,\Delta} |$. In summary, our goal is to determine a joint replenishment policy $T = (T_1, \ldots, T_n)$ that minimizes long-run average operating costs, represented by
\[ F(T) ~~=~~ J(T) + \sum_{i \in [n]} C_i( T_i ) \ . \]

\paragraph{The joint ordering term $\bs{J(T)}$.} When one comes across representation~\eqref{eqn:joint_cost_def} of the joint ordering term, it is only natural to ask why $\lim_{\Delta \to \infty} \frac{ N(T,\Delta) }{ \Delta }$ necessarily exists for any given policy $T$. To clarify this point, for any subset of commodities ${\cal N} \subseteq [n]$, let $M_{\cal N}$ be the least common multiple of $\{ T_i \}_{i \in {\cal N}}$, with the agreement that $M_{\cal N} = \infty$ when these time intervals do not have common multiples. Lemma~\ref{lem:limit_joint_order} below states a well-known observation, showing that the limit in question indeed exists and providing an explicit inclusion-exclusion-like formula; for a complete proof, we refer the reader to \citep[Sec.~1.1]{Segev23JRP}. That said, since the resulting expression consists of $2^n$ terms, its current form does not allow us to efficiently compute the joint ordering cost of arbitrarily-structured policies, implying that our algorithmic developments would have to circumvent this obstacle.

\begin{lemma} \label{lem:limit_joint_order}
$\lim_{\Delta \to \infty} \frac{ N(T,\Delta) }{ \Delta } = \sum_{{\cal N} \subseteq [n]} \frac{ (-1)^{ |{\cal N}| + 1 } }{ M_{\cal N} }$.
\end{lemma}

\subsection{Known results and open questions} \label{subsec:related_work}

Given the immense body of work dedicated to studying joint replenishment problems, including rigorous methods, heuristics, experimental evidence, industrial applications, and software solutions, there is no way to exhaustively review this literature. Hence, in what follows, we will be discussing only state-of-the-art results, directly pertaining to our research questions. For an in-depth literature review, readers are referred to excellent survey articles \citep{AksoyE88, GoyalS89, MuckstadtR93, KhoujaG08, BastosMNMC17} and book chapters \citep{SilverP85, SimchiLeviCB, Zipkin00, MuckstadtS10}, as well as to the references therein. Moreover, to learn about exciting progress with respect to joint replenishment in discrete time, one could consult selected papers along these lines \citep{LeviRSS08, BienkowskiBCJNS14, GayonMRS17, BosmanO20, SuriyanarayanaSGS24}.

\paragraph{The variable-base setting: Cornerstone algorithmic methods.} To the best of our knowledge, the now-classical works of \citet{Roundy85, Roundy86}, \cite{JacksonMM85}, \cite{MaxwellM85}, and \cite{MuckstadtR87} were the first to devise efficient and provably-good policies for the joint replenishment problem. At a high level, these papers proposed innovative ways to exploit natural convex relaxations, rounding their optimal solutions to so-called power-of-$2$ policies. Within the latter class, one fixes a common base, say $T_{\min}$, with each time interval $T_i$ being of the form $2^{ q_i } \cdot T_{\min}$, for some integer $q_i \geq 0$. Let us first examine the variable-base joint replenishment setting, where $T_1, \ldots, T_n$ are allowed to take arbitrary values. Quite amazingly, this mechanism for synchronizing joint orders can be employed to optimize $T_{\min}$ and $\{q_i\}_{i \in [n]}$, ending up with a power-of-$2$ policy whose long-run average cost is within factor $\frac{ 1 }{ \sqrt{2} \ln 2 } \approx 1.02$ of optimal! Since then, these findings have become some of the most renowned breakthroughs in inventory management, due to their widespread applicability, both theoretically and in practice. 

With the above-mentioned approximation guarantee standing for nearly four decades, our recent work \citep{Segev23JRP} finally attained the long-awaited improvement, showing that optimal policies can be efficiently approximated within any degree of accuracy. Technically speaking, this paper developed a new algorithmic approach for addressing the variable-base model, termed $\Psi$-pairwise alignment, enabling us to determine a replenishment policy whose long-run average cost is within factor $1 + \eps$ of optimal. For any $\eps \in (0, \frac{ 1 }{ 2 })$, the running time of our algorithm is $O( 2^{ \tilde{O}(1/\eps^3) } \cdot n^{ O(1) } )$, corresponding to the notion of an efficient polynomial-time approximation scheme (EPTAS). 

Unfortunately, both the design of such policies and their analysis are very involved, mostly due to utilizing an exponentially-large value for the alignment parameter, $\Psi = 2^{ \poly(1/\eps) }$. The latter choice, which seems unavoidable in light of our cost-bounding arguments, concurrently leads to an $\tilde{O}( \frac{ 1 }{ \eps^3 })$ running time exponent. Consequently, the primary open questions that motivate Section~\ref{sec:EPTAS-variable-base} of the present paper can be briefly summarized as follows:
\begin{quote}
{\em 
\begin{itemize}
    \item Can we come up with simpler arguments for analyzing $\Psi$-pairwise alignment?

    \item Is this method sufficiently accurate with $\Psi = \poly(1/\eps)$? 
    
    \item If doable, what are the running time implications of such improvements?
\end{itemize}}
\end{quote}

\paragraph{The fixed-base setting: Best known algorithmic methods.} Now, let us shift our attention to the fixed-base joint replenishment setting, where the time intervals $T_1, \ldots, T_n$ are restricted to being integer multiples of a prespecified time unit, $\Delta$. As one discovers by delving into previously-mentioned surveys \citep{AksoyE88, GoyalS89, MuckstadtR93, KhoujaG08, BastosMNMC17}, the latter requirement is motivated by real-life applications, where for practical concerns, cycle times are expected to be days, weeks, or months; here, policies with $T_i = \sqrt{e}$ or $T_i = \ln 7$, for example, cannot be reasonably implemented.

Yet another great achievement of \citet{Roundy85, Roundy86}, \cite{JacksonMM85}, and subsequent authors resides in proposing an ingenious rounding method for efficiently identifying feasible power-of-$2$ policies whose long-run average cost is within factor $\sqrt{9/8} \approx 1.06$ of optimal! Here, to ensure feasibility, rather than allowing the common base $T_{\min}$ to take arbitrary values, one should optimize this parameter over integer multiples of the basic unit, $\Delta$, explaining why approaching optimal policies in this model appears to be more challenging in comparison to its variable-base counterpart. While $\sqrt{9/8} \approx 1.06$ is widely believed to be the best approximation guarantee in this context, it has actually been surpassed years ago by \citet[Sec.~4]{TeoB01}. Their work developed a very elegant randomized rounding method for approximating the fixed-base joint replenishment problem within a factor of roughly $1.043$.

Unfortunately, due to a host of technical difficulties, we still do not know whether the notion of $\Psi$-pairwise alignment, in any conceivable form, can be adapted to provide improved approximation guarantees for the fixed-base setting. Given this state of affairs, the fundamental questions driving Section~\ref{sec:fixed_base_approx} of the present work, as stated in countless papers, books, conference talks, and course materials, can be concisely recapped as follows:
\begin{quote}
{\em 
\begin{itemize}
    \item Can we outperform the above-mentioned results, in any shape or form?

    \item It is easy to see why approximation guarantees for the fixed-base model readily migrate to the variable-base case; what about the opposite direction?
\end{itemize}}
\end{quote}

\paragraph{Integer-ratio policies and Roundy's conjecture.} Somewhat informally, the term ``integer-ratio'' has been coined by classical literature to capture families of replenishment policies $(T_1, \ldots, T_n)$ for which there exists some $T_0 > 0$, such that the ratio $\frac{ T_i }{ T_0 }$ is either an integer or its reciprocal, for every commodity $i \in [n]$. To qualify the slight informality, we mention that one could ask $\frac{ T_1 }{ T_0 }, \ldots, \frac{ T_n }{ T_0 }$ to reside within some proper subset $R$ of the positive integers and their reciprocals, in which case we arrive at $R$-integer-ratio policies or at additional twists on this terminology. 

Needless to say, power-of-$2$ policies are high-structured special cases of integer-ratio policies, since we are fixing a common base $T_{\min}$ and restricting each time interval $T_i$ to take the form $2^{ q_i } \cdot T_{\min}$, for some integer $q_i \geq 0$. From this perspective, one could view existing $\frac{ 1 }{ \sqrt{2} \ln 2 }$-approximations for the variable-base joint replenishment problem as utilizing a very specific type of integer-ratio policies. Interestingly, to this day, $\frac{ 1 }{ \sqrt{2} \ln 2 }$ constitutes the best-known factor achievable through integer-ratio policies, and it is entirely unclear whether the latter family offers strictly stronger performance guarantees. To motivate Section~\ref{sec:integer_ratio_evenly} of this paper, we take the following quote from the concluding remarks of Roundy's seminal work \citeyearpar[Sec.~8]{Roundy85}, stating a well-known conjecture about how good could integer-ratio policies potentially be:
\begin{quote}
``{\em It is likely that by using a somewhat more general class of policies, such as allowing $r_n = 2/3$ or $r_n = 3/2$, or by finding an optimal integer-ratio policy, the worst-case effectiveness of the lot-sizing rules given herein could be improved}''. 
\end{quote}

\paragraph{Incorporating resource constraints.} A common thread passing across all settings reviewed up until now is that different commodities are interacting ``only'' via their joint orders. Last but not least, we will be considering the resource-constrained joint replenishment problem, where the underlying commodities are further connected through the following family of constraints:
\[ \sum_{i \in [n]} \frac{ \alpha_{id} }{ T_i } ~~\leq~~ \beta_d \qquad \qquad \forall \, d \in [D] \]
While serving a wide range of practical purposes, it is instructive to take a production-oriented view on this model, where assembling each commodity $i$ requires $D$ limited resources. In turn, deciding on a time interval of $T_i$ between successive orders translates to consuming $\frac{ \alpha_{id} }{ T_i }$ units of each resource $d \in [D]$, whose overall capacity is denoted by $\beta_d$. For an elaborate discussion on the theoretical and practical usefulness of such constraints and their rich history, one could consult the classical work of \cite{Dobson87}, \cite{JacksonMM88}, and \cite{Roundy89} along these lines. 

Unfortunately, resource constraints appear to be rendering the joint replenishment problem significantly more difficult to handle in comparison to its unrestricted counterpart. To our knowledge, \cite{Roundy89} still holds the best approximation guarantee in this context, showing that sophisticated scaling arguments lead to the design of power-of-$2$ policies whose long-run average cost is within factor $\frac{ 1 }{ \ln 2 } \approx 1.442$ of optimal. Interestingly, subject to a single resource constraint, \cite{Roundy89} proved that the latter approximation guarantee can be improved to $\sqrt{9/8} \approx 1.06$, and we refer readers to the work of \cite{TeoB01} for very elegant methods to derive these results via randomized rounding. Given that the above-mentioned findings have been state-of-the-art for about 3.5 decades, the open questions that will lie at the heart of Sections~\ref{sec:RC_general} and~\ref{sec:RC_constant} can be succinctly highlighted as follows:
\begin{quote}
{\em 
\begin{itemize}
    \item For the general problem setup, can we devise improved rounding methods, perhaps deviating from power-of-$2$ policies?
    
    \item For a single resource constraint, could $\Psi$-pairwise alignment be exploited to breach the $\sqrt{9/8}$-barrier?
\end{itemize}}
\end{quote}

\paragraph{Hardness results.} Even though our work is algorithmically driven, to better understand the overall landscape, it is worth briefly mentioning known intractability results. Along these lines, \cite{SchulzT11} were the first to rigorously investigate how plausible it is to efficiently compute optimal replenishment policies in continuous time. Specifically, they proved that in the fixed-base setting, a polynomial time algorithm for the joint replenishment problem would imply an analogous result for integer factorization, thereby unraveling well-hidden connections between this question and fundamental problems in number theory. Subsequently, following Zhang's extraordinary work \citeyearpar{Zhang14} on bounded gaps between successive primes, \cite{CohenHillelY18} attained traditional complexity-based results, proving that the fixed-base setting is in fact strongly NP-hard. The latter result has been substantially simplified by \cite{SchulzT22}, showing that NP-hardness arises even in the presence of only two commodities. Finally, for the variable-base setting, \cite{SchulzT22} extended their original findings to derive its polynomial-relatability to integer factorization. The latter result was further lifted to a strong NP-hardness proof by \cite{TuisovY20}.

\subsection{Main contributions} \label{subsec:contributions}

The primary contribution of this paper resides in developing a wide range of algorithmic methods and analytical ideas --- some being completely novel and some offering enhancements to well-known techniques --- for resolving all open questions listed in Section~\ref{subsec:related_work}. For most of these questions, our results constitute the first quantitative improvement over power-of-$2$ policies and their nearby derivatives, which have been state-of-the-art in terms of provable performance guarantees since the mid-80's. In what follows, we provide a formal description of our main findings, leaving their structural characterization, algorithmic techniques, and analytical arguments to be discussed in subsequent sections. As an aside, while the next few paragraphs are titled ``Main result $\langle \text{number} \rangle$: \ldots'', this order is uncorrelated with importance; rather, it simply corresponds to the logical presentation order of these results.

\paragraph{Main result 1: The variable-base setting.} In Section~\ref{sec:EPTAS-variable-base}, we propose a new approach to streamline $\Psi$-pairwise alignment, simplifying the design principles of this framework and sharpening its performance guarantees. Interestingly, rather than utilizing an exponentially-large alignment parameter, our analysis will reveal that $\Psi = \tilde{O}( \frac{ 1 }{ \eps^3 } )$ suffices to identify $(1-\eps)$-approximate replenishment policies, while concurrently improving the running time exponent from $\tilde{O}( \frac{ 1 }{ \eps^3 } )$ to $\tilde{O}( \frac{ 1 }{ \eps } )$. The outcome of this analysis can be formalized as follows.

\begin{theorem} \label{thm:new_EPTAS}
For any $\eps > 0$, the variable-base joint replenishment problem can be approximated within factor $1 + \eps$ of optimal. The running time of our algorithm is $O( 2^{ \tilde{O}(1/\eps) } \cdot n^{ O(1) } )$.
\end{theorem}

\paragraph{Main result 2: The fixed-base setting.} In Section~\ref{sec:fixed_base_approx}, we describe a black-box reduction, showing that any approximation guarantee with respect to the variable-base convex relaxation \citep{Roundy85, Roundy86, JacksonMM85} readily migrates to the fixed-base joint replenishment problem, while incurring negligible loss in optimality. As an immediate implication, it follows that the fixed-base model can be efficiently approximated within factor $\frac{ 1 }{ \sqrt{2} \ln 2 } + \eps$. The specifics of this result, which surpasses the long-standing performance guarantees of $1.043$ due to \citet{TeoB01}, can be briefly stated as follows.  

\begin{theorem} \label{thm:fixed_base_main}
For any $\eps > 0$, the fixed-base joint replenishment problem can be approximated within factor $\frac{ 1 }{ \sqrt{2} \ln 2 } + \eps$ of optimal. The running time of our algorithm is $O( 2^{ O(1/\eps^2) } \cdot n^{ O(1) } )$.
\end{theorem}

\paragraph{Main result 3: Integer-ratio policies and Roundy's conjecture.} In Section~\ref{sec:integer_ratio_evenly}, we resolve Roundy's conjecture in the affirmative, improving on the best-known approximation factor achievable through integer-ratio policies, $\frac{ 1 }{ \sqrt{2} \ln 2 } \approx 1.02014$. Quite surprisingly, evenly-spaced policies will be shown to offer strictly stronger performance guarantees in comparison to their power-of-$2$ counterparts. In the former class of policies, we decide in advance to place evenly-spaced joint orders, and subsequently set all time intervals $T_1, \ldots, T_n$ as integer multiples of our spacing parameter, $\Delta$, which is a decision variable. 

\begin{theorem} \label{thm:even_spaced_better}
Optimal evenly-spaced policies approximate the joint replenishment problem within factor of at most $1.01915$.
\end{theorem}

It is important to emphasize that, for ease of presentation, we have not fully optimized the above-mentioned constant, which should primarily be viewed as a proof of concept. Possible avenues toward improvements in this context  are discussed in Section~\ref{sec:conclusions}. From a computational perspective, we will explain how to efficiently compute an evenly-spaced policy whose long-run average cost is within factor $1 + \eps$ of the optimal such policy. From a practical standpoint, we expect evenly-spaced policies to be particularly appealing, mainly due to their simplicity and implementability.

\paragraph{Main result 4: Incorporating resource constraints.} In Section~\ref{sec:RC_general}, we revisit a well-known convex relaxation of the resource-constrained joint replenishment problem, thinking about how optimal solutions can be better converted into feasible policies. By fusing together classical ideas and new structural insights, we devise a randomized rounding procedure that breaches the best-known approximation factor in this setting, $\frac{ 1 }{ \ln 2 } \approx 1.442$ \citep{Roundy89}. Once again, for simplicity of presentation, we avoid making concentrated efforts to minimize the resulting constant.

\begin{theorem} \label{thm:approx_JRP_general_RC}
The resource-constrained joint replenishment problem can be approximated in polynomial time within factor $1.417$ of optimal.
\end{theorem}

Finally, in  Section~\ref{sec:RC_constant}, we study the extent to which performance guarantees can be stretched, when running times are allowed to be exponential in the number of resource constraints, $D$. Along these lines, we propose an $O( n^{\tilde{O}( D^3/\eps^4 )} )$-time enumeration-based approach for developing a linear relaxation, arguing that its optimal fractional solution can be rounded into a near-optimal resource-feasible policy. The specifics of this result can be succinctly summarized  as follows.

\begin{theorem} \label{thm:approx_JRP_O1_RC}
For any $\eps > 0$, the resource-constrained joint replenishment problem can be approximated within factor $1 + \eps$ of optimal. The running time of our algorithm is $O( n^{\tilde{O}( D^3/\eps^4 )} )$. 
\end{theorem}

One consequence of the latter finding is that, when $D = O(1)$, we actually obtain a polynomial-time approximation scheme (PTAS). In particular, by setting $D=1$, this outcome improves on the classic $\sqrt{9/8}$-approximation for a single resource constraint \citep{Roundy89, TeoB01}.

%% file: TEX-EPTAS-Variable-Base.tex
\section{The Variable-Base Model: Improved Approximation Scheme} \label{sec:EPTAS-variable-base}

This section is dedicated to establishing Theorem~\ref{thm:new_EPTAS}, arguing that the variable-base joint replenishment problem can be approximated within factor $1 + \eps$ of optimal in $O( 2^{ \tilde{O}(1/\eps) } \cdot n^{ O(1) } )$ time. To this end, Sections~\ref{subsec:alg_definitions} and~\ref{subsec:LB_num_starR} introduce the basics of $\Psi$-pairwise alignment and reveal its newly-discovered lower-bounding method. Sections~\ref{subsec:alg_prelim} and~\ref{subsec:construct_policy} present a high-level overview of our algorithmic approach, followed by a deeper dive into its finer details. The performance guarantees of this approach are analyzed in Sections~\ref{subsec:cost_joint_orders} and~\ref{subsec:cost_comm_orders}.

\subsection{The primitives of \bstitle{\Psi}-pairwise alignment} \label{subsec:alg_definitions}

In what follows, we bring the reader up to speed, by introducing the basic ingredients of $\Psi$-pairwise alignment. While some of the upcoming definitions follow the overall approach of our previous work \citep{Segev23JRP}, others are  fundamentally different, and their intended role will be fleshed out along the way. As a side note, all notions discussed below are only serving analytical purposes, meaning that one should not be concerned with efficient implementation or with unknown pieces of information. These algorithmic questions will be addressed in subsequent sections. 

\paragraph{Interval classification.}  Let us make use of $T^* = (T_1^*, \ldots, T_n^*)$ to denote an optimal replenishment policy, fixed from this point on, with $T_{\min}^* = \min_{i \in [n]} T_i^*$ being the minimal time interval of any commodity. Given an error parameter $\eps \in (0, \frac{ 1 }{ 2 })$, we classify each interval $T_i^*$ as being large when $T_i^* > \frac{ 1 }{ \eps } \cdot T^*_{\min}$. In the opposite scenario,  $T_i^* \in [T^*_{\min}, \frac{ 1 }{ \eps } \cdot T^*_{\min}]$, in which case this interval will be referred to as being small. For a refined treatment of the latter class, we geometrically partition $[T^*_{\min}, \frac{ 1 }{ \eps } \cdot T^*_{\min}]$ by powers of $1 + \eps$, to obtain the sequence of segments $S_1^*, \ldots, S_L^*$. Specifically,    
\begin{equation} \label{eqn:def_segments}
S_1^* ~~=~~ [T^*_{\min}, (1 + \eps) \cdot T^*_{\min}), \quad S_2^* ~~=~~ [(1 + \eps) \cdot T^*_{\min}, (1 + \eps)^2 \cdot T^*_{\min}), \quad \ldots   
\end{equation}
so on and so forth, where in general $S_{\ell}^* = [(1 + \eps)^{\ell-1} \cdot T^*_{\min}, (1 + \eps)^{\ell} \cdot T^*_{\min})$. Here, $L$ is the minimal integer $\ell$ for which $(1 + \eps)^{\ell} \geq \frac{ 1 }{ \eps }$, meaning that $L = \lceil \log_{1+\eps} (\frac{ 1 }{ \eps } )\rceil \leq \frac{ 2 }{ \eps } \ln \frac{ 1 }{ \eps }$.

\paragraph{Active segments and  representatives.} We say that the segment $S_{\ell}^*$ is active when there is at least one commodity $i \in [n]$ with $T_i^* \in S_{\ell}^*$, letting ${\cal A}^* \subseteq [L]$ be the index set of active segments. Next, for each active segment $S_{\ell}^*$, let $R^*_{\ell}$ be an arbitrarily picked interval $T_i^*$ that belong to this segment; $R^*_{\ell}$ will be called the representative of $S_{\ell}^*$. 
 
We proceed by listing two useful observations regarding the set of representatives ${\cal R}^* = \{ R^*_{\ell} \}_{ \ell \in {\cal A}^* }$. First, Observation~\ref{obs:rel_NRstar_NTstar} informs us that, for every $\Delta \geq 0$, the number of joint orders across $[0,\Delta]$ with respect to the time intervals ${\cal R}^*$ is upper-bounded by the analogous quantity with respect to the optimal policy $T^*$; this claim can be straightforwardly inferred by noting that ${\cal R}^* \subseteq \{ T_1^*, \ldots, T_n^* \}$. Second, Observation~\ref{obs:Rstar1_in_Rstar} indirectly states that $S_1^*$ must be an active segment, meaning that its representative $R_1^*$ belongs to ${\cal R}^*$. To verify this claim, it suffices to note that $T_{\min}^* \in S_1^*$, by definition~\eqref{eqn:def_segments} of this segment.

\begin{observation} \label{obs:rel_NRstar_NTstar}
$N( {\cal R}^*, \Delta ) \leq N( T^*, \Delta )$, for every $\Delta \geq 0$.
\end{observation}

\begin{observation} \label{obs:Rstar1_in_Rstar}
$R_1^* \in {\cal R}^*$.
\end{observation}

\paragraph{$\bs{\Psi}$-pairwise alignment.} We say that a pair of active segments $S_{\ell_1}^*$ and $S_{\ell_2}^*$ is aligned when their representatives $R^*_{\ell_1}$ and $R^*_{\ell_2}$ have common integer multiples, which is equivalent to $\frac{ R^*_{\ell_1} }{ R^*_{\ell_2} }$ being a rational number. Moving on to define a stronger requirement, letting $M_{\ell_1, \ell_2}^*$ be the least common multiple of $R^*_{\ell_1}$ and $R^*_{\ell_2}$, this pair of segments is called $\Psi$-aligned when the corresponding multiples $\frac{ M_{\ell_1, \ell_2}^* }{ R^*_{\ell_1} }$ and $\frac{ M_{\ell_1, \ell_2}^* }{ R^*_{\ell_2} }$ both take values of at most $\Psi$, with the latter constant set to $\Psi = \frac{ 2 }{ \eps^3 } \ln^2 (\frac{ 1 }{ \eps })$. In this case, we make use of $\alpha^*_{ \{ \ell_1, \ell_2 \}, \ell_1 }$ and $\alpha^*_{ \{ \ell_1, \ell_2 \}, \ell_2 }$ to denote these two multiples, respectively. It is worth pointing out that, since $\Psi$ is chosen to be polynomial in $\frac{ 1 }{ \eps }$, we will have to come up with cost-bounding arguments that are significantly different from those of our recent work \citep{Segev23JRP}, which are relevant only when $\Psi = 2^{ \poly(1 / \eps ) }$.

\paragraph{The alignment graph.} Letting ${\cal P}_{\Psi}^*$ be the collection of $\Psi$-aligned pairs, in subsequent sections we will be exploiting the so-called alignment graph, $G_{\Psi}^* = ({\cal A}^*, {\cal P}_{\Psi}^*)$. Namely, the vertex set of this graph is comprised of the active segments, and each pair of such segments is connected by an edge when they are $\Psi$-aligned. We make use of
${\cal C}_1^*, \ldots, {\cal C}_{\Lambda}^*$ to denote the underlying connected components of $G_{\Psi}^*$. 

\subsection{The relation between  \bstitle{N( {\cal R}^*, \Delta )} and \bstitle{G_{\Psi}^*}} \label{subsec:LB_num_starR}

Let us be reminded that $N({\cal R}^*,\Delta)$ stands for the number of joint orders across $[0,\Delta]$ with respect to the set of representatives ${\cal R}^* = \{ R^*_{\ell} \}_{ \ell \in {\cal A}^* }$. One particular crux of our revised analysis consists in arguing that, up to $\eps$-dependent terms, $N({\cal R}^*,\Delta)$ is determined by the relationship between pairs of representatives within each connected component of $G_{\Psi}^*$. Moreover, these components will be contributing toward $N({\cal R}^*,\Delta)$ in a completely additive way. 

To formalize this statement, for every active segment $S_\ell^*$, let ${\cal M}_{R^*_{\ell},\Delta} = \{ 0, R^*_{\ell}, 2R^*_{\ell}, \ldots, \lfloor \frac{ \Delta }{ R^*_{\ell} } \rfloor \cdot R^*_{\ell} \}$ be the integer multiples of $R^*_{\ell}$ within $[0,\Delta]$. As such, the number of joint orders $N({\cal R}^*,\Delta)$ can be written as
\[ N({\cal R}^*,\Delta) ~~=~~ \left| \bigcup_{\ell \in {\cal A}^*} {\cal M}_{R^*_{\ell},\Delta} \right| ~~=~~ \left| \bigcup_{\lambda \in [\Lambda]} \bigcup_{\ell \in {\cal C}_{ \lambda }^*} {\cal M}_{R^*_{\ell},\Delta} \right| \ , \] 
and by the union bound, we clearly have
\[ N({\cal R}^*,\Delta) ~~\leq~~ \sum_{\lambda \in [\Lambda]} \left| \bigcup_{\ell \in {\cal C}_{ \lambda }^*} {\cal M}_{R^*_{\ell},\Delta} \right| ~~=~~ \sum_{\lambda \in [\Lambda]} N( {\cal R}^{*(\lambda)}, \Delta) \ , \]
with the convention that ${\cal R}^{*(\lambda)}$ stands for the set of representatives belonging to component ${\cal C}_{\lambda}^*$, i.e., ${\cal R}^{*(\lambda)} = \{ {\cal R}^*_{\ell} \}_{ \ell \in {\cal C}_{\lambda}^* }$. However, our crucial finding is that these terms can also be related in the opposite direction, arguing that $\sum_{\lambda \in [\Lambda]} N( {\cal R}^{*(\lambda)}, \Delta)$ nearly matches $N({\cal R}^*,\Delta)$, up to an additive error that depends only on $| {\cal A}^* |$. 

\begin{lemma} \label{lem:lower_N_U_new}
$(1 - \eps) \cdot \sum_{\lambda \in [\Lambda]} N( {\cal R}^{*(\lambda)}, \Delta) - \left| {\cal A}^* \right|^2 \leq N({\cal R}^*,\Delta) \leq \sum_{\lambda \in [\Lambda]} N( {\cal R}^{*(\lambda)}, \Delta)$.
\end{lemma}
\begin{proof}
To establish the desired lower bound, we make use of Bonferroni's inequality \citeyearpar{bonferroni1936} to obtain
\begin{eqnarray}
N({\cal R}^*,\Delta) & = & \left| \bigcup_{\lambda \in [\Lambda]} \bigcup_{\ell \in {\cal C}_{ \lambda }^*} {\cal M}_{R^*_{\ell},\Delta} \right| \nonumber \\
& \geq & \sum_{\lambda \in [\Lambda]} N( {\cal R}^{*(\lambda)}, \Delta) - \sum_{ \MyAbove{ \lambda_1, \lambda_2 \in [\Lambda]: }{ \lambda_1 < \lambda_2 } } \left| \left( \bigcup_{\ell \in {\cal C}^*_{\lambda_1}} {\cal M}_{R^*_{\ell},\Delta} \right) \cap \left( \bigcup_{\ell \in {\cal C}_{\lambda_2}^*} {\cal M}_{R^*_{\ell},\Delta} \right) \right| \ . \label{eqn:by_bonferroni}       
\end{eqnarray}
Focusing on a single pair of components $\lambda_1 \neq \lambda_2$, we proceed by examining the question of how large could their corresponding term be. For readability purposes, the proof of this claim appears immediately following the current one (see page~\pageref{page:proof_clm_bound_crossing_comp}).

\begin{claim} \label{clm:bound_crossing_comp}
$| ( \bigcup_{\ell \in {\cal C}^*_{\lambda_1}} {\cal M}_{R^*_{\ell},\Delta} ) \cap ( \bigcup_{\ell \in {\cal C}_{\lambda_2}^*} {\cal M}_{R^*_{\ell},\Delta} ) | \leq | {\cal C}_{\lambda_1}^* | \cdot | {\cal C}_{\lambda_2}^* | \cdot (  \frac{  N({\cal R}^*,\Delta) }{ \Psi } + 1 )$.
\end{claim}

Plugging this bound back into inequality~\eqref{eqn:by_bonferroni}, we have \begin{eqnarray}
N({\cal R}^*,\Delta) & \geq &  \sum_{\lambda \in [\Lambda]} N( {\cal R}^{*(\lambda)}, \Delta) - \left(  \frac{  N({\cal R}^*,\Delta) }{ \Psi } + 1 \right) \cdot \sum_{ \MyAbove{ \lambda_1, \lambda_2 \in [\Lambda]: }{ \lambda_1 < \lambda_2 } }  | {\cal C}_{\lambda_1}^* | \cdot | {\cal C}_{\lambda_2}^* | \nonumber \\
& \geq &  \sum_{\lambda \in [\Lambda]} N( {\cal R}^{*(\lambda)}, \Delta) - \left(  \frac{  N({\cal R}^*,\Delta) }{ \Psi } + 1 \right) \cdot \frac{ \left| {\cal A}^* \right|^2 }{ 2 } \label{eqn:lem_lower_N_U_1} \\
& \geq &  \sum_{\lambda \in [\Lambda]} N( {\cal R}^{*(\lambda)}, \Delta) - \eps \cdot N({\cal R}^*,\Delta) -  \left| {\cal A}^* \right|^2 \ , \label{eqn:lem_lower_N_U_2} 
\end{eqnarray}
and by rearranging, it indeed follows that $N({\cal R}^*,\Delta) \geq (1 - \eps) \cdot  \sum_{\lambda \in [\Lambda]} N( {\cal R}^{*(\lambda)}, \Delta) - \left| {\cal A}^* \right|^2$. Here, inequality~\eqref{eqn:lem_lower_N_U_2} holds since $\Psi = \frac{ 2 }{ \eps^3 } \ln^2 (\frac{ 1 }{ \eps })$, as stated in Section~\ref{subsec:alg_definitions}, and since $| {\cal A}^* | \leq L \leq \frac{ 2 }{ \eps } \ln \frac{ 1 }{ \eps }$, as argued in Section~\ref{subsec:alg_definitions}. The trickier transition is inequality~\eqref{eqn:lem_lower_N_U_1}, where we upper-bound $\sum_{ \lambda_1, \lambda_2 \in [\Lambda]: \lambda_1 < \lambda_2 } | {\cal C}_{\lambda_1}^* | \cdot | {\cal C}_{\lambda_2}^* |$ as follows: By observing that $\bigcup_{\lambda \in [\Lambda]} | {\cal C}_{\lambda}^* |$ is precisely the number of active segments, $|{\cal A}^*|$, we drive the desired bound via the next continuous relaxation:
\begin{equation} \label{eqn:opt_prob_crossing} \tag{P} \begin{array}{lll}
\max & {\displaystyle \sum_{ \MyAbove{ \lambda_1, \lambda_2 \in [\Lambda]: }{ \lambda_1 < \lambda_2 } }  x_{\lambda_1} x_{\lambda_2} } \\
\text{s.t.} & \| x \|_1 = |{\cal A}^*| \\
& x \in \bbR^{ \Lambda }_+
\end{array}
\end{equation}
In Appendix~\ref{app:proof_clm_Exact_opt_prob_crossing}, we prove the following claim, implying that $\sum_{ \lambda_1, \lambda_2 \in [\Lambda]: \lambda_1 < \lambda_2 } | {\cal C}_{\lambda_1}^* | \cdot | {\cal C}_{\lambda_2}^* | \leq \frac{ |{\cal A}^*|^2 }{ 2 }$.

\begin{claim} \label{clm:Exact_opt_prob_crossing}
$\opt\eqref{eqn:opt_prob_crossing} = \frac{ \Lambda - 1 }{ 2 \Lambda } \cdot |{\cal A}^*|^2$.
\end{claim}
\end{proof}

\paragraph{Proof of Claim~\ref{clm:bound_crossing_comp}.} \label{page:proof_clm_bound_crossing_comp} For any pair of connected components ${\cal C}_{\lambda_1}^* \neq {\cal C}_{\lambda_2}^*$, we obtain the upper bound in question by observing that 
\begin{eqnarray}
\left| \left( \bigcup_{\ell \in {\cal C}^*_{\lambda_1}} {\cal M}_{R^*_{\ell},\Delta} \right) \cap \left( \bigcup_{\ell \in {\cal C}_{\lambda_2}^*} {\cal M}_{R^*_{\ell},\Delta} \right) \right| & \leq & \sum_{ \ell_1 \in {\cal C}_{\lambda_1}^* } \sum_{ \ell_2 \in {\cal C}_{\lambda_2}^* } \left| {\cal M}_{R^*_{\ell_1},\Delta} \cap {\cal M}_{R^*_{\ell_2},\Delta} \right| \nonumber \\
& \leq &  \sum_{ \ell_1 \in {\cal C}_{\lambda_1}^* } \sum_{ \ell_2 \in {\cal C}_{\lambda_2}^* } \left( \left\lfloor \frac{ \Delta }{ \Psi \cdot \min \{ R^*_{\ell_1}, R^*_{\ell_2} \} } \right\rfloor + 1 \right) \label{eqn:clm_bound_crossing_comp_1} \\
& \leq &  | {\cal C}_{\lambda_1}^* | \cdot | {\cal C}_{\lambda_2}^* | \cdot \left(  \frac{ \Delta }{ \Psi \cdot T^*_{\min} } + 1 \right) \label{eqn:clm_bound_crossing_comp_2} \\
& \leq &  | {\cal C}_{\lambda_1}^* | \cdot | {\cal C}_{\lambda_2}^* | \cdot \left(  \frac{  N({\cal R}^*,\Delta) }{ \Psi } + 1 \right)  \ . \label{eqn:clm_bound_crossing_comp_3} 
\end{eqnarray}
To better understand where inequality~\eqref{eqn:clm_bound_crossing_comp_1} is coming from, the important observation is that, for any pair of segments $\ell_1$ and $\ell_2$ in different connected components of $G_{\Psi}^*$, we know that $R^*_{ \ell_1 }$ and $R^*_{ \ell_2 }$ are not $\Psi$-aligned. Namely, $R^*_{ \ell_1 }$ and $R^*_{ \ell_2 }$ either do not have common integer multiples, or have their least common multiple being greater than $\Psi \cdot \min \{ R^*_{\ell_1}, R^*_{\ell_2} \}$. Inequality~\eqref{eqn:clm_bound_crossing_comp_2} holds since both $R^*_{\ell_1}$ and $R^*_{\ell_2}$ take values of at least $T^*_{\min}$. Finally, inequality~\eqref{eqn:clm_bound_crossing_comp_3} is obtained by noting that $N({\cal R}^*,\Delta) \geq \frac{ \Delta }{ T_{\min}^* }$.

\subsection{Algorithmic preliminaries} \label{subsec:alg_prelim}

Having laid down the foundations of $\Psi$-pairwise alignment, we proceed with a distinction between two regimes --- one very easy to handle, and the other requiring our full-blown machinery. For this purpose, we remind the reader that $T^* = (T_1^*, \ldots, T_n^*)$ stands for an optimal replenishment policy, with $T_{\min}^* = \min_{i \in [n]} T_i^*$ being the minimal time interval of any commodity. We begin by computing an over-estimate $\widetilde{\opt}$ for the optimal long-run average cost $F(T^*)$, such that $F(T^*) \leq \widetilde{\opt} \leq 2 \cdot F(T^*)$. One way to obtain such an estimate in polynomial time is by computing a $\frac{ 1 }{ \sqrt{2} \ln2 }$-approximate power-of-$2$ policy, as explained in Section~\ref{subsec:related_work}. To avoid cumbersome notation, we plug in an approximation guarantee of $2$ rather than $\frac{ 1 }{ \sqrt{2} \ln2 } \approx 1.02$, noting that the specific constant does not play an important role.

\paragraph{The cheap ordering regime: $\bs{J(T^*) \leq \eps^2 \widetilde{\opt}}$.} Starting with the easy scenario, we argue that when the optimal joint ordering cost $J(T^*)$ is sufficiently small in comparison to $\widetilde{\opt}$, a rather straightforward replenishment policy is near-optimal. To this end, our candidate policy $\tilde{T} = (\tilde{T}_1, \ldots, \tilde{T}_n)$ is determined as follows:
\begin{itemize}
    \item {\em Placing joint orders:} We place joint order at integer multiples of $\Delta = \frac{ K_0 }{ \eps \widetilde{\opt} }$.

    \item {\em Placing commodity-specific orders:} For each commodity $i \in [n]$, let $T_i^{ \eoq }$ be the optimal solution to the standard EOQ model of this commodity (see Section~\ref{subsec:model_definition}). Namely, $T_i^{ \eoq }$ minimizes the long-run average cost $C_i( T_i ) = \frac{ K_i }{ T_i } + H_i T_i$, implying that $T_i^{ \eoq } = \sqrt{ K_i / H_i }$ by Claim~\ref{clm:EOQ_properties}. Given these definitions, we set the time interval of commodity $i$ as $\tilde{T}_i =\lceil T_i^{ \eoq } \rceil^{  (\Delta) } $, where $\lceil \cdot \rceil^{ (\Delta) }$ is an operator that rounds its argument up to the nearest integer multiple of $\Delta$.    
\end{itemize}
The next claim shows that, in the currently considered regime, this policy happens to be near-optimal. 

\begin{lemma} \label{lem:cheap_order_regime}
When ${J(T^*) \leq \eps^2 \widetilde{\opt}}$, we have $F(\tilde{T}) \leq (1 + 3\eps) \cdot F(T^*)$.
\end{lemma}
\begin{proof}
Recalling that $F(\tilde{T}) = J(\tilde{T}) + \sum_{i \in [n]} C_i( \tilde{T}_i )$, we proceed by separately bounding these two terms, showing that $J(\tilde{T}) \leq 2\eps \cdot F(T^*)$ and $\sum_{i \in [n]} C_i( \tilde{T}_i ) \leq (1+\eps) \cdot F(T^*)$. First, for the long-run joint ordering cost, since joint orders are placed at integer multiples of $\Delta = \frac{ K_0 }{ \eps \widetilde{\opt} }$, we have
\[ J(\tilde{T}) ~~=~~ \frac{ K_0 }{ \Delta } ~~=~~ \eps \widetilde{\opt} ~~\leq~~ 2\eps \cdot F(T^*) \ . \]
Moving on to consider marginal EOQ-based costs, note that
\begin{eqnarray}
\sum_{i \in [n]} C_i( \tilde{T}_i ) & = & \sum_{i \in [n]} \left( \frac{ K_i }{ \tilde{T}_i } + H_i \tilde{T}_i \right) \nonumber \\
& = & \sum_{i \in [n]} \left( \frac{ K_i }{ \lceil T_i^{ \eoq } \rceil^{  (\Delta) } } + H_i \cdot \lceil T_i^{ \eoq } \rceil^{  (\Delta) } \right) \nonumber \\
& \leq & \sum_{i \in [n]} \left( \frac{ K_i }{  T_i^{ \eoq } } + H_i \cdot \left( T_i^{ \eoq } + \Delta \right) \right) \nonumber \\
& = & \sum_{i \in [n]} C_i( T_i^{ \eoq } ) + \Delta \cdot \sum_{i \in [n]} H_i \nonumber \\
& \leq & \sum_{i \in [n]} C_i( T_i^* ) + \eps \cdot \sum_{i \in [n]} H_i T^*_i \label{eqn:lem_cheap_order_regime_1} \\
& \leq & (1+\eps) \cdot F(T^*) \ .\nonumber
\end{eqnarray} 
Here, inequality~\eqref{eqn:lem_cheap_order_regime_1} holds since $T_i^{ \eoq }$ minimizes $C_i( \cdot )$, implying in particular that $C_i( T_i^{ \eoq } ) \leq C_i( T_i^* )$. In addition, $\frac{ K_0 }{ T^*_{\min} } \leq J( T^* ) \leq \eps^2 \widetilde{\opt}$ by the case hypothesis, and therefore, $\Delta = \frac{ K_0 }{ \eps \widetilde{\opt} } \leq \eps  T^*_{\min} \leq \eps T^*_i$.
\end{proof}

\paragraph{The expensive ordering regime: $\bs{J(T^*) > \eps^2 \widetilde{\opt}}$.} We have now landed at the difficult scenario, where the vast majority of algorithmic effort is required. In Sections~\ref{subsec:construct_policy}-\ref{subsec:cost_comm_orders}, our objective is to efficiently construct a set of representative points $\tilde{\cal R} \subseteq \bbR_+$ that ``mimics'' the unknown set of optimal representatives ${\cal R}^* = \{ R^*_{\ell} \}_{ \ell \in {\cal A}^* }$, in the sense of simultaneously being $\eps$-dense and $\eps$-assignable. To better understand these properties, it is instructive to keep in mind the following interpretation:
\begin{enumerate}
\item {\em One-to-one correspondence:} This notion means that our set of representatives can be written as $\tilde{\cal R} = \{ \tilde{R}_{\ell} \}_{ \ell \in {\cal A}^* }$. We mention in passing that, while each optimal representative $R^*_{\ell}$ resides within $S^*_{\ell}$, its analogous $\tilde{R}_{\ell}$ will be allowed to slightly exceed this segment.

\item {\em $\eps$-density:} The set $\tilde{\cal R}$ is called $\eps$-dense when, by placing joints orders at all integer multiples of all representative points, we obtain an ordering density $\lim_{\Delta \to \infty} \frac{ N(\tilde{\cal R},\Delta) }{ \Delta }$  that matches the analogous density $\lim_{\Delta \to \infty} \frac{ N(T^*,\Delta) }{ \Delta }$ with respect to the optimal policy $T^*$, up to a factor of $1 + \eps$. By representation~\eqref{eqn:joint_cost_def}, this property translates to $J( \tilde{\cal R} ) \leq (1 + \eps) \cdot J( T^* )$, implying that our long-run joint ordering cost is near-optimal.

\item {\em $\eps$-assignability:} We say that $\tilde{\cal R}$ is $\eps$-assignable when, for each commodity $i \in [n]$, we can choose an integer multiple of some representative in $\tilde{\cal R}$ to serve as the time interval $\tilde{T}_i$ of this commodity, such that its marginal operating cost $C_i( \tilde{T}_i )$ is within factor $1 + \eps$ of the analogous cost $C_i( T_i^* )$ with respect to $T^*$.
\end{enumerate} 

\subsection{Constructing our replenishment policy} \label{subsec:construct_policy}

\paragraph{Step 1: Estimating $\bs{T^*_{\min}}$.} Let us first observe that, since $\frac{ 1 }{ T^*_{\min} }$ forms an upper bound on the ordering frequency of each commodity, we have $J(T^*) \leq \frac{ n K_0 }{ T_{\min}^* }$. Combining the latter inequality with our case hypothesis in the expensive ordering regime, $J(T^*) > \eps^2  \widetilde{\opt}$, it follows that $T_{\min}^* \leq \frac{ n }{ \eps^2 } \cdot \frac{K_0 }{ \widetilde{\opt} }$. On the other hand, $\frac{ K_0 }{ T_{\min}^* } \leq J(T^*) < F(T^*) \leq  \widetilde{\opt}$, meaning that $T_{\min}^* \geq \frac{  K_0 }{ \widetilde{\opt} }$. As such, we know that the minimal time interval $T_{\min}^*$ resides within $[\frac{ K_0 }{ \widetilde{\opt} }, \frac{ n }{ \eps^2 } \cdot \frac{ K_0 }{ \widetilde{\opt} })$. By enumerating over $O( \frac{ 1 }{ \eps } \log \frac{ n }{ \eps } )$ candidate values, this property allows us to assume that we have at our possession an under-estimate $\tilde{T}_{\min}$ of the minimal time interval $T_{\min}^*$, specifically, one that satisfies
\begin{equation} \label{eqn:rel_tildeTmin_Tmin}
( 1 - \eps ) \cdot T_{\min}^* ~~\leq~~ \tilde{T}_{\min} ~~\leq~~ T_{\min}^* \ . 
\end{equation}

\paragraph{Step 2: Guessing active segments.} Recalling that ${\cal A}^* \subseteq [L]$ stands for the index set of active segments, this set is clearly unknown from an algorithmic perspective. Therefore, our next step consists of guessing the precise identity of ${\cal A}^*$, or equivalently, whether each of the segments $S_1^*, \ldots, S_L^*$ is active or not. For this purpose, the overall number of guesses to consider is $2^L = O( 2^{ O( \frac{ 1 }{ \eps } \log \frac{ 1 }{ \eps } ) } )$.

\paragraph{Step 3: Guessing a spanning forest.} As explained in Section~\ref{subsec:alg_definitions}, the alignment graph $G_{\Psi}^* = ({\cal A}^*, {\cal P}_{\Psi}^*)$ is a  useful way to view the collection of $\Psi$-aligned pairs and to study their relationships. In this graph, our vertex set is comprised of the active segments ${\cal A}^*$, which is already known following step~2. However, noting that $G_{\Psi}^*$ connects each pair of such segments by an edge when they are $\Psi$-aligned, we are still unaware of how this edge set ${\cal P}_{\Psi}^*$ is structured. Moving forward, we will not be guessing the precise identity of ${\cal P}_{\Psi}^*$, which would require us to consider all $2^{ \Omega( |{\cal A}^*|^2 ) }$ possible subsets of edges, and in turn, to exceed the $O( 2^{ \tilde{O}(1/\eps) } \cdot n^{ O(1) } )$ running time guarantee stated in Theorem~\ref{thm:new_EPTAS}.

Instead, let us remind the reader that the underlying connected components of $G_{\Psi}^*$ were designated by ${\cal C}_1^*, \ldots, {\cal C}_{\Lambda}^*$. Letting ${\cal T}^*_{ \lambda }$ be an arbitrary spanning tree of each such component ${\cal C}_{\lambda }^*$, we proceed by guessing the entire forest ${\cal F}^* = \{ {\cal T}^*_1, \ldots, {\cal T}^*_{\Lambda} \}$. To this end, it suffices to enumerate across all possible forests over the set of vertices ${\cal A}^*$, where by Cayley's formula (see, e.g., \citet[pg.~235-240]{AignerZ18}), there are only $| {\cal A}^* |^{ O( | {\cal A}^* | )} = O( 2^{ O( \frac{ 1 }{ \eps } \log^2 (\frac{ 1 }{ \eps }) ) } )$ forests to consider.

\paragraph{Step 4: Guessing edge multiples.} Noting that ${\cal F}^*$ is a spanning forest of the alignment graph $G_{\Psi}^*$, we know that for each edge $(\ell_1, \ell_2) \in {\cal F}^*$, its corresponding pair $(R^*_{\ell_1}, R^*_{\ell_2})$ of optimal representatives is $\Psi$-aligned. In other words, letting $M_{\ell_1, \ell_2}^*$ be the least common multiple of $R^*_{\ell_1}$ and $R^*_{\ell_2}$, the corresponding multiples $\alpha^*_{ \{ \ell_1, \ell_2 \}, \ell_1 } = \frac{ M_{\ell_1, \ell_2}^* }{ R^*_{\ell_1} }$ and $\alpha^*_{ \{ \ell_1, \ell_2 \}, \ell_2 } = \frac{ M_{\ell_1, \ell_2}^* }{ R^*_{\ell_2} }$ both take values of at most $\Psi = \frac{ 2 }{ \eps^3 } \ln^2 (\frac{ 1 }{ \eps })$. Consequently, we can guess the latter multiples for all edges in ${\cal F}^*$ by enumerating over $O( \Psi^{ O(|E({\cal F}^*)|) } ) = O( 2^{ O( \frac{ 1 }{ \eps } \log^2 (\frac{ 1 }{ \eps }) ) } )$ options.

\paragraph{Step 5: Defining approximate representatives.} Given these ingredients, our revised method for defining the set of approximate representatives $\tilde{\cal R} = \{ \tilde{R}_{\ell} \}_{ \ell \in {\cal A}^* }$ makes completely independent decisions for each of the trees ${\cal T}^*_1, \ldots, {\cal T}^*_{\Lambda}$. Specifically, focusing on a single tree ${\cal T}^*_{ \lambda }$, let $\sigma_{ \lambda }$ be an arbitrarily picked vertex in ${\cal T}^*_{ \lambda }$, to which we refer as the source of this tree. Recalling that $R^*_{\sigma_{ \lambda }}$ is the representative of $S_{\sigma_{ \lambda }}^* = [(1 + \eps)^{\sigma_{ \lambda }-1} \cdot T^*_{\min}, (1 + \eps)^{\sigma_{ \lambda }} \cdot T^*_{\min})$, we begin by setting $\tilde{R}_{\sigma_{ \lambda }} = (1 + \eps)^{\sigma_{ \lambda }} \cdot \tilde{T}_{\min}$, which corresponds to the right endpoint of this segment, with the unknown $T^*_{\min}$ replaced by its estimate, $\tilde{T}_{\min}$. As a side note, any choice of $\tilde{R}_{\sigma_{ \lambda }}$ that can be $(1 \pm O(\eps))$-scaled back into $S_{\sigma_{ \lambda }}^*$ will be good enough for our purposes; choosing $\tilde{R}_{\sigma_{ \lambda }}$ as a proxy for the right endpoint is mainly for notational convenience.  

The important observation is that, once we fix a particular value for a single representative in ${\cal T}^*_{ \lambda }$, all other representatives in this tree are uniquely determined through the multiples $\{ (\alpha^*_{ \{ \ell_1, \ell_2 \}, \ell_1 }, \alpha^*_{ \{ \ell_1, \ell_2 \}, \ell_2 }) : (\ell_1, \ell_2) \in {\cal T}^*_{ \lambda } \}$. Indeed, let us consider some vertex $\ell \in {\cal T}^*_{ \lambda }$, with $\sigma_{ \lambda } = u_1, \ldots, u_k = \ell$ being the sequence of vertices along the unique $\sigma_{ \lambda }$-$\ell$ path in ${\cal T}^*_{ \lambda }$.  We first observe that since $(u_1, u_2) \in {\cal T}^*_{ \lambda } \subseteq G_{\Psi}^*$, to instill the exact same $\Psi$-alignment between $\tilde{R}_{u_1}$ and $\tilde{R}_{u_2}$, one should enforce $\alpha^*_{ \{ u_1, u_2 \}, u_1 } \cdot \tilde{R}_{u_1} = \alpha^*_{ \{ u_1, u_2 \}, u_2 } \cdot \tilde{R}_{u_2}$ for this particular pair. Similarly, since
$(u_2, u_3)$ is an edge of $G_{\Psi}^*$, this constraint sets $\alpha^*_{ \{ u_2, u_3 \}, u_2 } \cdot \tilde{R}_{u_2} = \alpha^*_{ \{ u_2, u_3 \}, u_3 } \cdot \tilde{R}_{u_3}$. Letting this relation propagate throughout the entire $\sigma_{ \lambda }$-$\ell$ path, its resulting sequence of equations can be aggregated to obtain a unique value for the representative $\tilde{R}_{\ell}$, given by:
\[ \tilde{R}_{\ell} ~~=~~ \tilde{R}_{u_k} ~~=~~ \left( \prod_{\kappa \in [k-1]} \frac{ \alpha^*_{ \{ u_{\kappa}, u_{\kappa+1} \}, u_{\kappa} } }{ \alpha^*_{ \{ u_{\kappa}, u_{\kappa+1} \}, u_{\kappa+1} } } \right) \cdot \tilde{R}_{u_1} ~~=~~ \left( \prod_{\kappa \in [k-1]} \frac{ \alpha^*_{ \{ u_{\kappa}, u_{\kappa+1} \}, u_{\kappa} } }{ \alpha^*_{ \{ u_{\kappa}, u_{\kappa+1} \}, u_{\kappa+1} } } \right) \cdot \tilde{R}_{\sigma_{\lambda}} \ . \]

\paragraph{Observation: Approximate representatives vs.\ optimal ones.} An important consequence of the preceding discussion is that our explanation of how a single representative determines its entire component applies to the  collection of optimal representatives $\{ R_{\ell}^* \}_{ \ell \in {\cal A}^* }$ as well. In particular, for every tree ${\cal T}^*_{ \lambda }$ and for every vertex $\ell \in {\cal T}^*_{ \lambda }$, we know that
\[ R^*_{\ell} ~~=~~ \left( \prod_{\kappa \in [k-1]} \frac{ \alpha^*_{ \{ u_{\kappa}, u_{\kappa+1} \}, u_{\kappa} } }{ \alpha^*_{ \{ u_{\kappa}, u_{\kappa+1} \}, u_{\kappa+1} } } \right) \cdot R^*_{\sigma_{\lambda}} \ , \]
where the constant above is identical to the one that relates between $\tilde{R}_{\ell}$ and $\tilde{R}_{\sigma_{\lambda}}$. An immediate conclusion is that, since $R^*_{\sigma_{\lambda}} \in S_{\sigma_{ \lambda }}^* = [(1 + \eps)^{\sigma_{ \lambda }-1} \cdot T^*_{\min}, (1 + \eps)^{\sigma_{ \lambda }} \cdot T^*_{\min})$ and since $\tilde{R}_{\sigma_{ \lambda }} = (1 + \eps)^{\sigma_{ \lambda }} \cdot \tilde{T}_{\min} \in [( 1 - \eps ) \cdot (1 + \eps)^{\sigma_{ \lambda }} \cdot T^*_{\min}, (1 + \eps)^{\sigma_{ \lambda }} \cdot T^*_{\min}]$ by equation~\eqref{eqn:rel_tildeTmin_Tmin}, the approximate representatives $\{ \tilde{R}_{\ell} \}_{ \ell \in {\cal T}^*_{ \lambda } }$ are proportional to their optimal counterparts $\{ R^*_{\ell} \}_{ \ell \in {\cal T}^*_{ \lambda } }$, up to a component-dependent multiplicative factor of $1 \pm \eps$, as formally stated below.

\begin{observation} \label{obs:scaling_component}
For every $\lambda \in [\Lambda]$, there exists a coefficient $\gamma_{\lambda} \in 1 \pm \eps$ such that $\tilde{R}_{\ell} = \gamma_{\lambda} \cdot R^*_{\ell}$ for every $\ell \in {\cal T}^*_{ \lambda }$.
\end{observation}

\paragraph{Step 6: The final policy.} We are now ready to lay down the specifics of our replenishment policy, which will be denoted by $\tilde{T} = (\tilde{T}_1, \ldots, \tilde{T}_n)$. To this end, given the set of approximate representatives $\tilde{\cal R} = \{ \tilde{R}_{\ell} \}_{ \ell \in {\cal A}^* }$, we proceed as follows:
\begin{itemize}
\item {\em Placing joint orders:} Joint orders will be placed only at integer multiples of the approximate representatives $\{ \tilde{R}_{\ell} \}_{ \ell \in {\cal A}^* }$. In other words, with ${\cal M}_{\tilde{R}_{\ell}} = \{ 0, \tilde{R}_{\ell}, 2\tilde{R}_{\ell}, \ldots \}$ standing for the integer multiples of $\tilde{R}_{\ell}$, we decide in advance to open a joint order at every point in $\bigcup_{ \ell \in {\cal A}^* } {\cal M}_{\tilde{R}_{\ell}}$, regardless of whether any given point will subsequently be utilized by some commodity or not.

\item {\em Placing commodity-specific orders:} For each commodity $i \in [n]$, we determine its time interval $\tilde{T}_i$ to be the one that minimizes its marginal EOQ cost $C_i(\cdot)$ out of the following options:
\begin{itemize}
\item {\em Small intervals:} Any of the approximate representatives $\{ \tilde{R}_{\ell} \}_{ \ell \in {\cal A}^* }$.

\item {\em Single large interval:} Letting $T^{\max}_i = \max \{ \frac{ 1 }{ \eps } \cdot \tilde{T}_{\min}, \sqrt{ K_i / H_i  } \}$, the additional option is $\lceil T^{\max}_i \rceil^{ (\tilde{R}_1) }$, where $\lceil \cdot \rceil^{ (\tilde{R}_1) }$ is an operator that rounds its argument up to the nearest integer multiple of $\tilde{R}_1$.
\end{itemize}
\end{itemize}
It is important to emphasize that, while choosing one of the ``small'' options as the time interval $\tilde{T}_i$ clearly falls within our set of joint orders, this also happens to be the case for the ``large''  option. Indeed, by Observation~\ref{obs:Rstar1_in_Rstar}, we know that $\tilde{R}_1 \in \tilde{\cal R}$, implying that ordering commodity $i$ according to the interval $\lceil T^{\max}_i \rceil^{ (\tilde{R}_1) }$ falls on integer multiples of $\tilde{R}_1$, where joint orders have already been placed.

\paragraph{Remark: Choosing a single policy.} It is imperative to point out that, since our algorithmic approach employs numerous guessing steps, to ultimately identify the least expensive policy out of all possible outcomes, one should be able to efficiently estimate the long-run cost function $F( \cdot )$ for each resulting policy $\tilde{T}$. However, while evaluating $\sum_{i \in [n]} C_i( \tilde{T}_i )$ is straightforward, a blind application of Lemma~\ref{lem:limit_joint_order} would lead to an exponentially-sized formula for the joint ordering cost $J( \tilde{T} )$.

To bypass this obstacle, let us first recall that each of our policies $\tilde{T}$ places joint orders only at integer multiples of its approximate representatives $\tilde{\cal R}$, implying that $J( \tilde{T} ) = J( \tilde{\cal R} )$. Now, one hidden feature of Sections~\ref{subsec:alg_definitions} and~\ref{subsec:LB_num_starR} is that their entire discussion holds for any replenishment policy, regardless of whether it is optimal or not. This observation brings us to conclude that Lemma~\ref{lem:lower_N_U_new} can be written in terms of $\tilde{T}$ rather than $T^*$, and therefore,
\[ J( \tilde{\cal R} ) ~~=~~ K_0 \cdot \lim_{ \Delta \to \infty } \frac{ N( \tilde{\cal R}, \Delta ) }{ \Delta } ~~\in~~ (1 \pm \eps) \cdot K_0 \cdot \sum_{\lambda \in [\Lambda]} \lim_{ \Delta \to \infty } \frac{ N( \tilde{\cal R}^{(\lambda)}, \Delta) }{ \Delta } \ . \]
In Appendix~\ref{app:proof_lem_eval_lim_U_Delta}, we explain how to evaluate the latter limit via an inclusion-exclusion formula in $O( 2^{ \tilde{O}(1/\eps) } )$ time. 

\begin{lemma} \label{lem:eval_lim_U_Delta}
$\lim_{ \Delta \to \infty } \frac{ N( \tilde{\cal R}^{(\lambda)}, \Delta) }{ \Delta }$ can be computed in $O( 2^{ \tilde{O}(1/\eps) } )$ time, for every $\lambda \in [\Lambda]$.
\end{lemma}

\subsection{Cost analysis: Joint orders} \label{subsec:cost_joint_orders}

Following the high-level outline of Section~\ref{subsec:alg_prelim}, our analysis begins by establishing $O(\eps)$-density. Recalling that the replenishment policy $\tilde{T}$ places joint orders at integer multiples of the approximate representatives $\tilde{\cal R} = \{ \tilde{R}_{\ell} \}_{ \ell \in {\cal A}^* }$, we argue that the latter set is $4\eps$-dense. In other words, we relate the ordering density of $\tilde{\cal R}$ to that of the optimal policy $T^*$, showing that 
\begin{equation} \label{eqn:lim_NhatT_NTstar}
\lim_{\Delta \to \infty} \frac{ N(\tilde{\cal R},\Delta) }{ \Delta } ~~\leq~~ (1 + 4\eps) \cdot \lim_{\Delta \to \infty} \frac{ N(T^*,\Delta) }{ \Delta } \ . 
\end{equation}
By representation~\eqref{eqn:joint_cost_def}, this property directly implies that our long-run joint ordering cost is near-optimal, in the sense that $J( \tilde{T} ) \leq (1 + 4\eps) \cdot J( T^* )$. To derive inequality~\eqref{eqn:lim_NhatT_NTstar}, it is worth mentioning that $N( {\cal R}^*, \Delta ) \leq N( T^*, \Delta )$ for every $\Delta \geq 0$, by Observation~\ref{obs:rel_NRstar_NTstar}. Therefore, the desired result will be a direct consequence of the next relation between $N(\tilde{\cal R},\Delta)$ and $N({\cal R}^*,\Delta)$.

\begin{lemma} \label{lem:UB_points_hatR}
$N(\tilde{\cal R},\Delta) \leq (1 + 4\eps) \cdot N({\cal R}^*,\Delta) + 4 \cdot | {\cal A}^* |^2$, for every $\Delta \geq 0$.
\end{lemma}
\begin{proof}
We remind the reader that $N(\tilde{\cal R},\Delta)$ stands for the number of joint orders across $[0,\Delta]$ with respect to the time intervals $\tilde{\cal R}$.  Letting ${\cal M}_{\tilde{R}_{\ell},\Delta} = \{ 0, \tilde{R}_{\ell}, 2\tilde{R}_{\ell}, \ldots, \lfloor \frac{ \Delta }{ \tilde{R}_{\ell} } \rfloor \cdot \tilde{R}_{\ell} \}$ be the integer multiples of $\tilde{R}_{\ell}$ within $[0,\Delta]$, we clearly have
\begin{eqnarray}
N(\tilde{\cal R},\Delta) & = & \left| \bigcup_{\ell \in {\cal A}^*} {\cal M}_{\tilde{R}_{\ell},\Delta} \right| \nonumber \\
& = & \left| \bigcup_{\lambda \in [\Lambda]} \bigcup_{\ell \in {\cal T}^*_{ \lambda }} {\cal M}_{\tilde{R}_{\ell},\Delta} \right| \nonumber \\
& \leq & \sum_{\lambda \in [\Lambda]} \left|  \bigcup_{\ell \in {\cal T}^*_{ \lambda }} {\cal M}_{\tilde{R}_{\ell},\Delta} \right| \ . \label{eqn:lem_UB_points_hatR_1}  
\end{eqnarray}
Here, the second equality is obtained by decomposing the set of active segments ${\cal A}^*$ according to the spanning forest ${\cal T}^*_1, \ldots, {\cal T}^*_{\Lambda}$, and the last inequality follows from the union bound.

The important observation is that, by Observation~\ref{obs:scaling_component}, for every tree ${\cal T}^*_{ \lambda }$ there exists a coefficient $\gamma_{\lambda} \in 1 \pm \eps$, such that each of the approximate representatives $\{ \tilde{R}_{\ell} \}_{ \ell \in {\cal T}^*_{ \lambda } }$ is a $\gamma_{\lambda}$-scaling of its optimal counterpart, i.e., $\tilde{R}_{\ell} = \gamma_{\lambda} \cdot R^*_{\ell}$. Therefore, the set of joint orders $\bigcup_{\ell \in {\cal T}^*_{ \lambda }} {\cal M}_{\tilde{R}_{\ell},\Delta}$ we are seeing in inequality~\eqref{eqn:lem_UB_points_hatR_1} can be viewed as a $\gamma_{\lambda}$-scaling of $\bigcup_{\ell \in {\cal T}^*_{ \lambda }} {\cal M}_{R^*_{\ell},\Delta/\gamma_{\lambda}}$. Consequently,
\begin{eqnarray*}
N(\tilde{\cal R},\Delta) & \leq & \sum_{\lambda \in [\Lambda]} \left|  \bigcup_{\ell \in {\cal T}^*_{ \lambda }} {\cal M}_{R^*_{\ell},\Delta/\gamma_{\lambda}}\right| \\
& \leq & (1 + \eps) \cdot \sum_{\lambda \in [\Lambda]} \left|  \bigcup_{\ell \in {\cal T}^*_{ \lambda }} {\cal M}_{R^*_{\ell},\Delta}\right| \\
& = & (1 + \eps) \cdot \sum_{\lambda \in [\Lambda]} N( {\cal R}^{*(\lambda)}, \Delta) \ .
\end{eqnarray*}
Combining this bound with the relation between $\sum_{\lambda \in [\Lambda]} N( {\cal R}^{*(\lambda)}, \Delta)$ and $N({\cal R}^*,\Delta)$ prescribed by Lemma~\ref{lem:lower_N_U_new}, namely $N({\cal R}^*,\Delta) \geq (1 - \eps) \cdot \sum_{\lambda \in [\Lambda]} N( {\cal R}^{*(\lambda)}, \Delta) - | {\cal A}^* |^2$, we conclude that $N(\tilde{\cal R},\Delta) \leq (1 + 4\eps) \cdot N({\cal R}^*,\Delta) + 4 \cdot | {\cal A}^* |^2$, as desired.
\end{proof}

\subsection{Cost analysis: Commodity-specific orders} \label{subsec:cost_comm_orders}

Our second analytical step consists of establishing $O(\eps)$-assignability, specifically showing that the set of representatives $\{ \tilde{R}_{\ell} \}_{ \ell \in {\cal A}^* }$ is $5\eps$-assignable. In other words, we argue that the marginal operating cost of each commodity with respect to our approximate replenishment policy $\tilde{T}$ is within factor $1 + 5\eps$ of the analogous quantity with respect to the optimal policy $T^*$. The precise nature of the latter relation can be formalized as follows.

\begin{lemma} \label{lem:UB_marginal_hatT}
$C_i( \tilde{T}_i ) \leq (1 + 5\eps) \cdot C_i( T_i^* )$, for every commodity $i \in [n]$.   
\end{lemma}
\begin{proof}
Our analysis is divided into three scenarios, depending on how the optimal solution $T_i^{ \eoq } = \sqrt{ K_i / H_i }$ to the standard EOQ model of each commodity relates to $T^*_{\min}$. As mentioned in Claim~\ref{clm:EOQ_properties}, $T_i^{ \eoq }$ is the unique minimizer of the long-run average cost $C_i( T_i ) = \frac{ K_i }{ T_i } + H_i T_i$. Specifically, we will be considering low, medium, and high regimes, where ``low'' corresponds to $T_i^{ \eoq } < T^*_{\min}$, ``medium'' represents $T_i^{ \eoq } \in [T_{\min}^*, \frac{ 1 }{ \eps } \cdot T_{\min}^*]$, and  ``high'' examines the residual case, $T_i^{ \eoq } > \frac{ 1 }{ \eps } \cdot T_{\min}^*$. For readability purposes, the low regime is discussed below, whereas the medium and high ones are deferred to Appendix~\ref{app:proof_lem_UB_marginal_hatT}.

\paragraph{The low regime: $\bs{T_i^{ \eoq } < T^*_{\min}}$.} In this case, let us circle back to Observation~\ref{obs:Rstar1_in_Rstar}, stating that $R_1^* \in {\cal R}^*$. Put differently, $S_1^*$ is necessarily an active segment, implying that the approximate set $\tilde{\cal R}$ constructed in Section~\ref{subsec:construct_policy} includes a representative of this segment, $\tilde{R}_1$. To tie between $\tilde{R}_1$ and $R_1^*$, let ${\cal C}_{\lambda}^*$ be the connected component of $G_{\Psi}^*$ where segment $S_1^*$ resides. Then, by Observation~\ref{obs:scaling_component}, we know that there exists a coefficient $\gamma_{\lambda} \in 1 \pm \eps$ such that $\tilde{R}_1 = \gamma_{\lambda} \cdot R^*_1$. Now, as explained in step~6 of  Section~\ref{subsec:construct_policy}, $\tilde{R}_1$ is one of the options considered for our time interval $\tilde{T}_i$. Since we pick the option that minimizes the EOQ cost $C_i(\cdot)$ of this commodity, 
\begin{eqnarray}
C_i( \tilde{T}_i ) & \leq & C_i( \tilde{R}_1 ) \nonumber \\
& \leq & (1 + 2\eps) \cdot C_i( R^*_1 ) \label{eqn:lem_UB_marginal_hatT_1} \\
& \leq & (1 + 5\eps) \cdot C_i( T^*_{\min} ) \label{eqn:lem_UB_marginal_hatT_2} \\
& \leq & (1 + 5\eps) \cdot C_i( T^*_i )  \ . \label{eqn:lem_UB_marginal_hatT_3}
\end{eqnarray}
Here, inequality~\eqref{eqn:lem_UB_marginal_hatT_1} holds since $\tilde{R}_1 = \gamma_{\lambda} \cdot R^*_1$, and it is easy verify that $C_i( \theta T ) \leq \max \{ \theta, \frac{ 1 }{ \theta } \} \cdot C_i(T)$ for all $\theta > 0$. Inequality~\eqref{eqn:lem_UB_marginal_hatT_2} follows from a similar argument, recalling that $R^*_1 \in S_1^* = [T^*_{\min}, (1 + \eps) \cdot T^*_{\min})$. Finally, to obtain inequality~\eqref{eqn:lem_UB_marginal_hatT_3}, note that since $C_i$ is a strictly convex function with 
a unique minimizer at $T_i^{ \eoq }$ (see Claim~\ref{clm:EOQ_properties}), it is strictly increasing over $[T_i^{ \eoq }, \infty)$. Therefore, we can indeed conclude that $C_i( T^*_{\min} ) \leq C_i( T^*_i )$ by observing that $T_i^{ \eoq } < T_{\min}^* \leq T_i^*$, where the first inequality is due to the case hypothesis of this regime.
\end{proof}

%% file: TEX-Black-Box-Reduction.tex
\section{Fixed-Base via Variable-Base: Black-Box Reduction} \label{sec:fixed_base_approx}

In what follows, we explain how any approximation guarantee with respect to the variable-base convex relaxation can essentially be migrated to the fixed-base joint replenishment problem. Consequently, as stated in Theorem~\ref{thm:fixed_base_main}, the fixed-base model will be shown to be approximable within factor $\frac{ 1 }{ \sqrt{2} \ln 2 } + \eps$ of optimal in $O( 2^{ O(1/\eps^2) } \cdot n^{ O(1) } )$ time. Moving forward, Section~\ref{subsec:overview_fixed_base} presents a high-level overview of our approach, leaving the finer details of its analysis to be discussed in Sections~\ref{subsec:proof_lem_black_box_marginal_1} and~\ref{subsec:proof_lem_black_box_bound_c2}.

\subsection{High-level overview} \label{subsec:overview_fixed_base}

With respect to a given fixed base $\Delta$, let $T^* = (T_1^*, \ldots, T_n^*)$ be an optimal replenishment policy in this context, and let $T^*_{\min} = \min_{i \in [n]} T_i^*$ be its corresponding minimal time interval. Our reduction proceeds by considering two cases, depending on the magnitude of $\rho^* = \frac{ T^*_{\min} }{ \Delta }$. In the fixed-base model, this parameter is obviously an integer.

\paragraph{Case 1: $\bs{\rho^* \leq \frac{ 1 }{ \eps }}$.} Let us assume without loss of generality that $\frac{ 1 }{ \eps }$ takes an integer value. As such, we first guess the exact value of $\rho^*$, for which there are only $\frac{ 1 }{ \eps }$ options by the case hypothesis. Consequently, $T^*_{\min} = \rho^* \Delta$ is known as well. Next, for each of the $O( \frac{ T^*_{\min } }{ \eps \Delta } )$ points $T^*_{\min}, T^*_{\min} + \Delta, T^*_{\min} + 2\Delta, \ldots, \frac{ 1 }{ \eps } \cdot T^*_{\min}$ we guess whether it is one of the time intervals $T^*_1, \ldots, T_n^*$ or not, letting ${\cal R}^*$ be the resulting set. Here, the total number of guesses is $2^{ O( \frac{ T^*_{\min } }{ \eps \Delta } )} = 2^{O( \rho^* / \eps )} = 2^{ O(1/\eps^2) }$. Given these guesses, our replenishment policy $\tilde{T}$ is constructed as follows:
\begin{itemize}
\item {\em Placing joint orders:} Joint orders will be placed only at integer multiples of ${\cal R}^*$. Clearly, since ${\cal R}^* \subseteq \{ T_1^*, \ldots, T_n^* \}$, it follows that our long-run joint ordering cost is $J( \tilde{T} ) \leq J( T^* )$. 

\item {\em Placing commodity-specific orders:} For each commodity $i \in [n]$, we determine its time interval $\tilde{T}_i$ to be the one that minimizes the EOQ cost $C_i(\cdot)$ out of the set ${\cal R}^* \cup \{ \bar{T}_i \}$. Here, $\bar{T}_i = \max \{ \frac{ 1 }{ \eps } \cdot T^*_{\min}, \lceil \sqrt{ K_i / H_i  } \rceil^{ \min } \}$, where $\lceil \cdot  \rceil^{ \min }$ is an operator that rounds its argument up to the nearest integer multiple of $T^*_{\min}$. It is important to emphasize that, by allowing only these options for choosing $\tilde{T}_i$, we are not creating new joint orders. In Section~\ref{subsec:proof_lem_black_box_marginal_1}, we prove the next claim, relating our EOQ-based costs to those of the optimal policy $T^*$.

\begin{lemma} \label{lem:black_box_marginal_1}
$C_i( \tilde{T}_i ) \leq (1 + \eps) \cdot C_i( T^*_i )$, for every commodity $i \in [n]$.    
\end{lemma}
\end{itemize}

\paragraph{Case 2: $\bs{\rho^* > \frac{ 1 }{ \eps }}$.} We have now landed at the scenario where existing results for the variable-base model will be useful. Noting that the current case hypothesis is equivalent to $T_{\min}^* > \frac{ \Delta }{ \eps }$, we proceed by plugging this constraint into the well-known convex relaxation of the variable-base model \citep{Roundy85, Roundy86, JacksonMM85}, to obtain:
\begin{equation} \label{eqn:conv_relax_large} \tag{$\mathrm{R}^+$} \begin{array}{lll}
\min & {\displaystyle \frac{ K_0 }{ T_{\min} } + \sum_{i \in [n]} \left( \frac{ K_i }{ T_i } + H_i T_i \right)} \qquad \\
\text{s.t.} & T_i \geq T_{\min} \geq \frac{ \Delta }{ \eps } & \forall \, i \in [n] 
\end{array}
\end{equation}
By employing deterministic power-of-$2$ rounding, as proposed by \citet[Sec.~2.2]{TeoB01}, we are guaranteed to construct a replenishment policy $\hat{T} = (\hat{T}_{\min}, \hat{T}_1, \ldots, \hat{T}_n)$ satisfying the next three properties:
\begin{enumerate}
    \item {\em Cost:} $F( \hat{T} ) \leq \frac{ 1 }{ \sqrt{2} \ln 2 } \cdot \opt\eqref{eqn:conv_relax_large} \leq \frac{ 1 }{ \sqrt{2} \ln 2 } \cdot F(T^*)$.
    
    \item {\em Minimal time interval:} $\hat{T}_{\min} = \min \{ \hat{T}_1, \ldots, \hat{T}_n \} \geq \frac{ 1 }{ \sqrt{2} } \cdot \frac{ \Delta }{ \eps }$.

    \item {\em Power-of-$2$ structure:} For every $i \in [n]$, the time interval $\hat{T}_i$ can be written as $2^{q_i} \cdot \hat{T}_{\min}$, for some integer $q_i \geq 0$.
\end{enumerate}
As a side note regarding property~2, expert readers can recall that \cite{TeoB01} scale every coordinate of an optimal solution to~\eqref{eqn:conv_relax_large} by at most $\sqrt{2}$ in either direction. Therefore, we indeed end up with $\hat{T}_i \geq \frac{ 1 }{ \sqrt{2} } \cdot \frac{ \Delta }{ \eps }$, due to incorporating the constraint $ T_i \geq T_{\min} \geq \frac{ \Delta }{ \eps }$ into this convex program.

Clearly, the fundamental issue with this policy is that $\hat{T}_1, \ldots, \hat{T}_n$ may not be integer multiples of the fixed base $\Delta$. To bypass this obstacle, note that $\hat{T}_i = 2^{q_i} \cdot \hat{T}_{\min}$, for some integer $q_i \geq 0$, by property~3. As such, we define a replenishment policy $\tilde{T}$ in which $\tilde{T}_i = 2^{q_i} \cdot \lceil \hat{T}_{\min} \rceil^{ (\Delta) }$ for every commodity $i \in [n]$, with the convention that $\lceil \cdot  \rceil^{ (\Delta) }$ is an operator that rounds its argument up to the nearest integer multiple of $\Delta$. In Section~\ref{subsec:proof_lem_black_box_bound_c2}, we prove the next result, showing that our combined operational cost matches the optimal one up to a factor of essentially $\frac{ 1 }{ \sqrt{2} \ln 2 }$.

\begin{lemma} \label{lem:black_box_bound_c2}
$F( \tilde{T} ) \leq (1+\sqrt{2} \eps) \cdot \frac{ 1 }{ \sqrt{2} \ln 2 } \cdot F(T^*)$. 
\end{lemma}

\subsection{Proof of Lemma~\ref{lem:black_box_marginal_1}} \label{subsec:proof_lem_black_box_marginal_1}

Our proof considers two cases, depending on the relation between $T_i^*$ and $\frac{ 1 }{ \eps } \cdot T^*_{\min}$. Starting with the straightforward case, when $T_i^* \leq \frac{ 1 }{ \eps } \cdot T^*_{\min}$, our guessing procedure guarantees that the time interval $T_i^*$ necessarily resides within ${\cal R}^*$. As such, one of the options considered for choosing $\tilde{T}_i$ is $T_i^*$, and therefore $C_i( \tilde{T}_i ) \leq C_i( T^*_i )$.

Now, when $T_i^* > \frac{ 1 }{ \eps } \cdot T^*_{\min}$, the important observation is that
\begin{equation} \label{eqn:lem_black_box_marginal_1_3}
C_i( T_i^* ) ~~\geq~~ \min \left\{ C_i(T) : T \geq \frac{ 1 }{ \eps } \cdot T^*_{\min} \right\} ~~=~~ C_i \left( \max \left\{ \frac{ 1 }{ \eps } \cdot T^*_{\min},  \sqrt{ \frac{ K_i }{ H_i }  } \right\} \right) \ ,  
\end{equation}
where the latter equality follows from Claim~\ref{clm:EOQ_properties}, stating in particular that $C_i$ is a strictly convex function, with a unique minimizer at $\sqrt{ K_i / H_i }$. On the other hand, one of the options considered for choosing $\tilde{T}_i$ is $\bar{T}_i$, and therefore,
\begin{eqnarray}
C_i( \tilde{T}_i ) & \leq & C_i( \bar{T}_i ) \nonumber \\
& = & C_i \left( \max \left\{ \frac{ 1 }{ \eps } \cdot T^*_{\min}, \left\lceil \sqrt{ \frac{ K_i }{ H_i }  } \right\rceil^{ \min } \right\} \right) \nonumber \\
& \leq & (1 + \eps) \cdot C_i \left( \max \left\{ \frac{ 1 }{ \eps } \cdot T^*_{\min},  \sqrt{ \frac{ K_i }{ H_i }  } \right\} \right) \label{eqn:lem_black_box_marginal_1_1} \\
& \leq & (1 + \eps) \cdot C_i( T_i^* )
\ . \label{eqn:lem_black_box_marginal_1_2}
\end{eqnarray}
Here, inequality~\eqref{eqn:lem_black_box_marginal_1_1} holds since $\lceil \sqrt{ K_i / H_i  } \rceil^{ \min } \leq \sqrt{ K_i / H_i  } + T^*_{\min}$, implying that
\[  \max \left\{ \frac{ 1 }{ \eps } \cdot T^*_{\min}, \left\lceil \sqrt{ \frac{ K_i }{ H_i }  } \right\rceil^{ \min } \right\} ~~\leq~~ (1+\eps) \cdot \max \left\{ \frac{ 1 }{ \eps } \cdot T^*_{\min}, \sqrt{ \frac{ K_i }{ H_i } } \right\} \ , \]
and one can easily verify that $C_i( \theta T ) \leq \theta \cdot C_i(T)$ for all $\theta \geq 1$. Inequality~\eqref{eqn:lem_black_box_marginal_1_2} is precisely the one obtained in~\eqref{eqn:lem_black_box_marginal_1_3}.

\subsection{Proof of Lemma~\ref{lem:black_box_bound_c2}} \label{subsec:proof_lem_black_box_bound_c2}

To account for the combined operational cost of $\tilde{T}$, we first observe that in terms of joint orders,
\[ J( \tilde{T} ) ~~=~~ \frac{ K_0 }{ \tilde{T}_{\min} } ~~=~~ \frac{ K_0 }{ \lceil \hat{T}_{\min} \rceil^{ (\Delta) } } ~~\leq~~ \frac{ K_0 }{ \hat{T}_{\min}  } ~~=~~ J( \hat{T} ) \ , \]
where the first and last equalities respectively hold since both $\tilde{T}$ and $\hat{T}$ are power-of-$2$ policies. Moving on to consider EOQ-based costs, we observe that $\lceil \hat{T}_{\min} \rceil^{ (\Delta) } \leq \hat{T}_{\min} + \Delta \leq (1 + \sqrt{2}\eps) \cdot \hat{T}_{\min}$, where the last inequality holds since $\hat{T}_{\min} \geq \frac{ 1 }{ \sqrt{2} } \cdot \frac{ \Delta }{ \eps }$, as mentioned in property~2. Consequently, for every commodity $i \in [n]$,
\[ C_i( \tilde{T}_i ) ~~=~~ C_i( 
2^{q_i} \cdot \lceil \hat{T}_{\min} \rceil^{ (\Delta) } ) ~~\leq~~ (1 + \sqrt{2}\eps) \cdot C_i( 2^{q_i} \cdot \hat{T}_{\min} ) ~~=~~ (1 + \sqrt{2}\eps) \cdot C_i( \hat{T}_i ) \ . \]
All in all, it follows that the long-run average cost of our  policy is
\begin{eqnarray*}
F( \tilde{T} ) & = & J( \tilde{T} ) + \sum_{i \in [n]} C_i( \tilde{T}_i ) \\
& \leq & J( \hat{T} ) + (1 + \sqrt{2}\eps) \cdot \sum_{i \in [n]}  C_i( \hat{T}_i ) \\
& \leq & (1 + \sqrt{2}\eps) \cdot F( \hat{T} ) \\
& \leq & (1 + \sqrt{2}\eps) \cdot \frac{ 1 }{ \sqrt{2} \ln 2 } \cdot F(T^*) \ ,
\end{eqnarray*}
where the last inequality holds since 
$F( \hat{T} ) \leq \frac{ 1 }{ \sqrt{2} \ln 2 } \cdot F(T^*)$, by property~1.

%% file: TEX-Evenly-Spaced.tex
\section{Evenly-Spaced Policies: Improved Guarantees for Integer-Ratio Policies} \label{sec:integer_ratio_evenly}

The main result of this section resides in constructively resolving Roundy's conjecture. As stated in Theorem~\ref{thm:even_spaced_better}, we prove that optimal evenly-spaced policies approximate the variable-base joint replenishment problem within factor of at most $1.01915$, thereby improving on the best-known guarantees achievable via integer-ratio policies. Toward this objective, Section~\ref{subsec:overview_evently_spaced} describes the main ideas of our proof, leaving most technical claims to be established in Sections~\ref{subsec:policies_even} and~\ref{subsec:proof_clm_bound_k0_vs_tmin}. It is important to emphasize that these contents form an existence proof, which is not algorithmic in nature. To address this issue, Section~\ref{subsec:approx_even_spaced} explains how to efficiently compute an evenly-spaced policy whose long-run average cost is within factor $1 + \eps$ of the optimal such policy. 

\subsection{High-level overview} \label{subsec:overview_evently_spaced}
 
Let $T^* = (T_1^*, \ldots, T_n^*)$ be an optimal replenishment policy, and let $T^*_{\min} = \min_{i \in [n]} T_i^*$ be its corresponding minimal time interval. We say that commodity $i$ is $T^*$-fractional when $T_i^*$ is not an integer multiple of $T^*_{\min}$. Clearly, when there are no $T^*$-fractional commodities, $T^*$ is already an evenly-spaced policy. In the opposite case, we make use of $f$ to denote the $T^*$-fractional commodity whose time interval is minimal. Our proof proceeds by considering two cases, depending on the relation between $T_f^*$ and $T^*_{\min}$.  
   
\paragraph{Case 1: $\bs{T_f^* > 3 T^*_{\min}}$.} \label{par:case1_even_space} Starting with the easier scenario, we define an evenly-spaced policy $\tilde{T}$ as follows:
\begin{itemize}
\item {\em Placing joint orders:} Joint orders will be placed at integer multiples of $T^*_{\min}$. As a result, $\tilde{T}$ is guaranteed to be an evenly-spaced policy, with a long-run ordering cost of $J(\tilde{T}) \leq J(T^*)$.

\item {\em Placing commodity-specific orders:} To determine the time interval $\tilde{T}_i$ of every commodity $i \in [n]$, our decision depends on how $T_i^*$ and $T^*_{\min}$ are related.
    \begin{itemize}
        \item {\em When $T_i^* \leq 3 T^*_{\min}$:} By the hypothesis of case~1, any such commodity is necessarily an integer multiple of $T^*_{\min}$, and we simply define $\tilde{T}_i = T^*_i$. Clearly, in terms of EOQ cost, $C_i( \tilde{T}_i ) = C_i( T^*_i )$.

        \item {\em When $T_i^* > 3 T^*_{\min}$:} For any such commodity, we have $T_i^* \in [\kappa_i, \kappa_i+1] \cdot T^*_{\min}$, for some integer $\kappa_i \geq 3$. Here, we set $\tilde{T}_i$ to be the better option out of $\kappa_i \cdot T^*_{\min}$ and $(\kappa_i+1) \cdot T^*_{\min}$ in terms of minimizing the marginal cost $C_i(\cdot)$. Consequently,
        \begin{eqnarray*}
        C_i( \tilde{T}_i ) & = & \min \left\{ C_i( \kappa_i \cdot T^*_{\min} ), C_i( (\kappa_i+1) \cdot T^*_{\min} ) \right\} \\
        & \leq & \frac{ 1 }{ 2 } \cdot \left( \sqrt{ \frac{ \kappa_i + 1 }{ \kappa_i } } + \sqrt{ \frac{ \kappa_i }{ \kappa_i+1 } } \right) \cdot C_i( T^*_i ) \\
        & \leq & \frac{ 1 }{ 2 } \cdot \left( \sqrt{ \frac{ 4 }{ 3 } } + \sqrt{ \frac{ 3 }{ 4 } } \right) \cdot C_i( T^*_i ) \\
        & \leq & 1.01037 \cdot C_i( T^*_i ) \ ,     
        \end{eqnarray*}
        where the first inequality follows from Claim~\ref{clm:EOQ_properties}(4), and the second inequality holds since the function $x \mapsto \sqrt{ \frac{ x + 1 }{ x } } + \sqrt{ \frac{ x }{ x+1 } }$ is decreasing over $(0,\infty)$.       
    \end{itemize} 
\end{itemize}

\paragraph{Case 2: $\bs{T_f^* < 3  T^*_{\min}}$.}
In this scenario, the crux of our argument would be to employ power-of-$2$ rounding directly on the optimal policy $T^*$, rather than with respect to an optimal solution to some convex relaxation. The first step in this direction consists of showing that, under the hypothesis of case~2, there is a meaningful gap between the optimal long-run joint ordering cost $J( T^* )$ and our long-run payment $\frac{ K_0 }{ T_{\min}^* }$ for orders containing the most frequent commodity. The next claim formalizes this connection, whose proof is deferred to Section~\ref{subsec:proof_clm_bound_k0_vs_tmin}.

\begin{claim} \label{clm:bound_k0_vs_tmin}
When $T_f^* < 3  T^*_{\min}$, we have $J(T^*) \geq \frac{ 6 }{ 5 } \cdot \frac{ K_0 }{ T_{\min}^* }$.
\end{claim}

Motivated by this result, we devise in Section~\ref{subsec:policies_even} two different ways of rounding $T^*$ into an evenly-spaced policy. In a nutshell, our first policy $\tilde{T}^A$ will make use of randomization to be particularly attractive when the joint ordering term $J(T^*)$ forms a large fraction of the optimal cost. Specifically, its expected joint ordering cost and EOQ-based cost will be designed to satisfy
\begin{equation} \label{eqn:even_policy_A}
\exsub{ \potwo }{ J( \tilde{T}^A ) } ~~\leq~~ \frac{ 1 }{ \sqrt{2} \ln 2 } \cdot \frac{ 5 }{ 6 } \cdot J( T^* ) \quad \text{and} \quad
\exsub{ \potwo }{ \sum_{i \in [n]} C_i( \tilde{T}^A_i ) } ~~=~~ \frac{ 1 }{ \sqrt{2} \ln 2 } \cdot \sum_{i \in [n]} C_i( T_i^* ) \ . 
\end{equation}
The second policy, $\tilde{T}^B$, will be appealing in the complementary scenario, when the EOQ-based cost $\sum_{i \in [n]} C_i( T_i^* )$ constitutes a large fraction. Here, we show how to end up with
\begin{equation} \label{eqn:even_policy_B}
 J( \tilde{T}^B ) ~~\leq~~ \frac{ 5 }{ 2 } \cdot J(T^*) \qquad \text{and} \qquad \sum_{i \in [n]} C_i( \tilde{T}^B_i ) ~~\leq~~ \frac{ 1 }{ 2 } \cdot \left( \sqrt{ \frac{ 4 }{ 3 } } + \sqrt{ \frac{ 3 }{ 4 } } \right) \cdot \sum_{i \in [n]} C_i( T^*_i ) \ . 
 \end{equation}

We proceed by arguing that a random selection between these policies guarantees a $1.01915$-approximation. To this end, our final policy $\tilde{T}$  picks $\tilde{T}^A$ with probability $\theta = 0.89755$ and $\tilde{T}^B$ with probability $1 - \theta$. As a result, 
\begin{eqnarray*}
\exsub{ \theta, \potwo }{ F( \tilde{T} ) } & = & \theta \cdot \exsub{ \potwo }{ F( \tilde{T}^A ) } + (1-\theta) \cdot F( \tilde{T}^B )  \\
& = & \theta \cdot \exsub{ \potwo }{ J( \tilde{T}^A ) } + (1-\theta) \cdot J( \tilde{T}^B ) \\
&& \mbox{} + \theta \cdot \exsub{ \potwo }{ \sum_{i \in [n]} C_i( \tilde{T}^A_i ) } + (1-\theta) \cdot \sum_{i \in [n]} C_i( \tilde{T}^B_i ) \\
& \leq & \left( \frac{ \theta }{ \sqrt{2} \ln 2 } \cdot \frac{ 5 }{ 6 } + (1 - \theta) \cdot \frac{ 5 }{ 2 } \right) \cdot  J(T^*) \\
&& \mbox{} + \left( \frac{ \theta }{ \sqrt{2} \ln 2 } + (1-\theta) \cdot \frac{ 1 }{ 2 } \cdot \left( \sqrt{ \frac{ 4 }{ 3 } } + \sqrt{ \frac{ 3 }{ 4 } } \right)  \right) \cdot C(T^*) \\
& \leq & 1.01915 \cdot (J(T^*) + C(T^*)) \\
& = & 1.01915 \cdot F(T^*) \ ,
\end{eqnarray*}
where the first inequality is obtained by plugging inequalities~\eqref{eqn:even_policy_A} and~\eqref{eqn:even_policy_B}.

\subsection{Policy design} \label{subsec:policies_even}

\paragraph{The policy $\bs{\tilde{T}^A}$.} Our first policy $\tilde{T}^A$ is obtained by employing power-of-$2$ rounding with respect to $T^*$, specifically by following the randomized approach of \citet[Sec.~2.2]{TeoB01}. Here, we are guaranteed to end up with a random replenishment policy $\tilde{T}^A = (\tilde{T}_1^A, \ldots, \tilde{T}_n^A)$ satisfying the next three properties:
\begin{enumerate}
    \item {\em Expectation:} $\exsubpar{\potwo}{ \tilde{T}^A_i } = \frac{ 1 }{ \sqrt{2} \ln 2 } \cdot T_i^*$ and $\exsubpar{\potwo}{ \frac{ 1 }{ \tilde{T}^A_i } } = \frac{ 1 }{ \sqrt{2} \ln 2 } \cdot \frac{ 1 }{ T_i^* }$, for every $i \in [n]$.

    \item {\em Relative order:} $\tilde{T}^A_{i_1} \leq \tilde{T}^A_{i_2}$ if and only if $T^*_{i_1} \leq T^*_{i_2}$, for every pair $i_1 \neq i_2$.

    \item {\em Power-of-$2$ structure:} For every $i \in [n]$, the time interval $\tilde{T}_i$ can be written as $2^{q_i} \cdot \tilde{T}_{\min}^A$, for some integer $q_i \geq 0$, where $\tilde{T}_{\min}^A = \min \{ \tilde{T}_1^A, \ldots, \tilde{T}_n^A \}$.
\end{enumerate}
As a result, our expected joint ordering cost is
\begin{eqnarray}
\exsub{\potwo}{ J( \tilde{T}^A ) } & = & \exsub{\potwo}{ \frac{ K_0 }{ \tilde{T}_{\min}^A } } \label{eqn:policy_TA_1} \\
& = & \frac{ 1 }{ \sqrt{2} \ln 2 } \cdot \frac{ K_0 }{ T^*_{\min} } \label{eqn:policy_TA_2}  \\
& \leq & \frac{ 1 }{ \sqrt{2} \ln 2 } \cdot \frac{ 5 }{ 6 } \cdot J( T^* ) \ . \label{eqn:policy_TA_3} 
\end{eqnarray}
Here, equality~\eqref{eqn:policy_TA_1} holds since, by property~3, all time intervals are integer multiples of $\tilde{T}_{\min}^A$. Inequality~\eqref{eqn:policy_TA_2} follows by combining properties~1 and~2. Finally, inequality~\eqref{eqn:policy_TA_3} is obtained by plugging $\frac{ K_0 }{ T_{\min}^* } \leq \frac{ 5 }{ 6 } \cdot J(T^*)$, as stated in Claim~\ref{clm:bound_k0_vs_tmin}. Moving on to examine the EOQ-based cost of our policy, we observe that
\begin{eqnarray}
\exsub{ \potwo }{ \sum_{i \in [n]} C_i( \tilde{T}^A_i ) } & = & \sum_{i \in [n]} \left( \exsub{ \potwo }{ \frac{ K_i }{ \tilde{T}^A_i } } + \exsub{ \potwo }{ H_i \tilde{T}^A_i } \right) \nonumber \\
& = & \frac{ 1 }{ \sqrt{2} \ln 2 } \cdot \sum_{i \in [n]} \left( \frac{ K_i }{ T^*_i } + H_i T^*_i \right) \nonumber \\
& = & \frac{ 1 }{ \sqrt{2} \ln 2 } \cdot \sum_{i \in [n]} C_i( T_i^* ) \ , \label{eqn:policy_TA_4}
\end{eqnarray}
where the second equality follows from property~1.

\paragraph{The policy $\bs{\tilde{T}^B}$.} The second policy we propose is deterministic, baring certain similarities to our approach in case~1 (see page~\pageref{par:case1_even_space}), albeit with an appropriate scaling of the ordering frequency. Specifically, the policy $\tilde{T}^B$ is structured as follows. 
\begin{itemize}
    \item {\em Placing joint orders:} Joint orders will be placed at integer multiples of $\Delta = \frac{ T^*_{\min} }{ 3 }$. As such, $\tilde{T}^B$ is an evenly-spaced policy, with a long-run ordering cost of 
    \begin{equation} \label{eqn:policy_TB_1}
    J( \tilde{T}^B ) ~~=~~ 3 \cdot \frac{ K_0 }{ T^*_{\min} } ~~\leq~~ \frac{ 5 }{ 2 } \cdot J(T^*) \ ,    
    \end{equation}
    where the last inequality follows from Claim~\ref{clm:bound_k0_vs_tmin}.

    \item {\em Placing commodity-specific orders:} Noting that $T^*_i \geq  T^*_{\min} =3 \Delta$ for every commodity $i \in [n]$, we have $T^*_i \in [\kappa_i, \kappa_i+1] \cdot \Delta$, for some integer $\kappa_i \geq 3$. This observation motives us to set $\tilde{T}_i^B$ as the better option out of $\kappa_i \cdot \Delta$ and $(\kappa_i+1) \cdot \Delta$ in terms of minimizing the marginal cost $C_i(\cdot)$. By Claim~\ref{clm:EOQ_properties}(4), it follows that
    \begin{eqnarray}
    C_i( \tilde{T}^B_i ) & = & \min \left\{ C_i( \kappa_i \cdot \Delta ), C_i( (\kappa_i+1) \cdot \Delta ) \right\} \nonumber \\
    & \leq & \frac{ 1 }{ 2 } \cdot \left( \sqrt{ \frac{ \kappa_i + 1 }{ \kappa_i } } + \sqrt{ \frac{ \kappa_i }{ \kappa_i+1 } } \right) \cdot C_i( T^*_i ) \nonumber \\
    & \leq & \frac{ 1 }{ 2 } \cdot \left( \sqrt{ \frac{ 4 }{ 3 } } + \sqrt{ \frac{ 3 }{ 4 } } \right) \cdot C_i( T^*_i ) \ . \label{eqn:policy_TB_2} 
    \end{eqnarray}
\end{itemize}

\subsection{Proof of Claim~\ref{clm:bound_k0_vs_tmin}}  \label{subsec:proof_clm_bound_k0_vs_tmin}

To derive the desired bound, $J(T^*) \geq \frac{ 6 }{ 5 } \cdot \frac{ K_0 }{ T_{\min}^* }$, it suffices to argue that $\lim_{\Delta \to \infty} \frac{ N(\{ T^*_{\min}, T_f^* \},\Delta) }{ \Delta } \geq \frac{ 6 }{ 5 } \cdot \frac{ 1 }{ T^*_{\min} }$, since 
$N(T^*,\Delta) \geq N(\{ T^*_{\min}, T_f^* \},\Delta)$ for any $\Delta \geq 0$. For this purpose, according to the inclusion-exclusion formula in Lemma~\ref{lem:limit_joint_order}, we have
\[ \lim_{\Delta \to \infty} \frac{ N(\{ T^*_{\min}, T_f^* \},\Delta) }{ \Delta } ~~=~~ \frac{ 1 }{ T^*_{\min} } + \frac{ 1 }{ T_f^* } - \frac{ 1 }{ \mylcm(T^*_{\min}, T_f^* ) } \ , \]
where $\mylcm(T^*_{\min}, T_f^* )$ stands for the least common multiple of $T^*_{\min}$ and $T_f^*$. Our proof proceeds by considering two cases, depending on the value of $\mylcm(T^*_{\min}, T_f^* )$. The latter quantity is either equal to $2T_f^*$ or at least $3T_f^*$, since $f$ is a $T^*$-fractional commodity.
\begin{itemize}
   \item {\em When $\mylcm(T^*_{\min}, T_f^* ) = 2T_f^*$:} Since $T_f^* < 3 T^*_{\min}$, we must have either $T_f^* = \frac{ 3 }{ 2 } \cdot T^*_{\min}$ or $T_f^* = \frac{ 5 }{ 2 } \cdot T^*_{\min}$. As a result, 
    \[ \frac{ 1 }{ T^*_{\min} } + \frac{ 1 }{ T_f^* } - \frac{ 1 }{ \mylcm(T^*_{\min}, T_f^* ) } ~~=~~ \frac{ 1 }{ T^*_{\min} } + \frac{ 1 }{ 2T_f^* } ~~\geq~~ \frac{ 6 }{ 5 } \cdot \frac{ 1 }{ T^*_{\min} } \ . \]
    
    \item {\em When $\mylcm(T^*_{\min}, T_f^* ) \geq 3T_f^*$:} Here,
    \[ \frac{ 1 }{ T^*_{\min} } + \frac{ 1 }{ T_f^* } - \frac{ 1 }{ \mylcm(T^*_{\min}, T_f^* ) } ~~\geq~~ \frac{ 1 }{ T^*_{\min} } + \frac{ 2 }{ 3T_f^* } ~~\geq~~ \frac{ 11 }{ 9 } \cdot \frac{ 1 }{ T^*_{\min} } \ , \]
    where the last inequality holds once again since $T_f^* < 3 T^*_{\min}$.
\end{itemize} 

\subsection{Approximating evenly-spaced policies} \label{subsec:approx_even_spaced}

In what follows, we describe the main ideas behind deriving an algorithmic version of Theorem~\ref{thm:even_spaced_better}, showing how to efficiently compute an evenly-spaced policy whose long-run average cost is within factor $1 + \eps$ of the optimal such policy. To avoid repetitive contents in relation to earlier sections, these ideas will only be sketched. 

\paragraph{Step 1: Enumerating over $\bs{\Delta}$.} Let $T^* = (T_1^*, \ldots, T_n^*)$ be an optimal evenly-spaced replenishment policy, and let $\Delta^*$ be its spacing parameter. Along the lines of Section~\ref{subsec:alg_prelim}, we begin by computing an over-estimate $\widetilde{\opt}$ for the optimal long-run average cost $F(T^*)$, such that $F(T^*) \leq \widetilde{\opt} \leq 2 \cdot F(T^*)$. The important observation is that there exists a $(1+4\eps)$-approximate evenly-spaced  policy, $\hat{T}$, whose spacing parameter $\hat{\Delta}$ resides within $[ \frac{ K_0 }{ \widetilde{\opt} }, \frac{ 1 }{ \eps } \cdot \frac{ K_0 }{ \widetilde{\opt} } ]$. Indeed, when $\Delta^*$ satisfies this property, $T^*$ is one such policy. Otherwise, since $\Delta^* \geq \frac{ K_0 }{ F(T^*) } \geq \frac{ K_0 }{ \widetilde{\opt} }$, the desired property may only be violated by having $\Delta^* >  \frac{ 1 }{ \eps } \cdot \frac{ K_0 }{ \widetilde{\opt} }$. In this case, we can keep the time intervals $T_1^*, \ldots, T_n^*$ unchanged, and simply scale down $\Delta^*$ by a factor of $\lceil \eps \Delta^* \cdot \frac{ \widetilde{\opt} }{ K_0 } \rceil$. With the latter modification, our joint ordering cost is still negligible, since 
\[ \frac{ K_0 }{ \Delta^* } \cdot \left\lceil \eps \Delta^* \cdot \frac{ \widetilde{\opt} }{ K_0 } \right\rceil ~~\leq~~ 2\eps \cdot \widetilde{\opt} ~~\leq~~ 4\eps \cdot F(T^*) \ . \]
As a result, by enumerating over $O( \frac{ 1 }{ \eps } \log \frac{ 1 }{ \eps } )$ candidate values, we can assume to have at our possession an over-estimate $\tilde{\Delta}$ of the spacing parameter $\hat{\Delta}$, such that $\hat{\Delta} \leq \tilde{\Delta} \leq (1 + \eps) \cdot \hat{\Delta}$.

\paragraph{Step 2: Placing commodity-specific orders.} We proceed by explaining why, once the spacing parameter $\Delta$ of an evenly-spaced policy has been fixed, optimally choosing its time intervals is a  straightforward task. To this end, for each commodity $i \in [n]$, let us recall that $T_i^{ \eoq }$ denotes the optimal solution to the standard EOQ model of this commodity (see Section~\ref{subsec:model_definition}). Namely, $T_i^{ \eoq }$ minimizes the long-run average cost $C_i( T_i ) = \frac{ K_i }{ T_i } + H_i T_i$, implying that $T_i^{ \eoq } = \sqrt{ K_i / H_i }$ by Claim~\ref{clm:EOQ_properties}. In parallel, the latter claim informs us that $C_i$ is strictly convex, meaning that when we are restricted to integer multiples of $\Delta$, the optimal choice for $T_i$ must be either $\lfloor T_i^{ \eoq } \rfloor^{  (\Delta) }$ or $\lceil T_i^{ \eoq } \rceil^{  (\Delta) }$, i.e., the nearest multiples of $\Delta$ from below and above, respectively. Consequently, we determine each time interval $\tilde{T}_i$ by choosing the $C_i$-cheaper option out of $\lfloor T_i^{ \eoq } \rfloor^{  (\tilde{\Delta}) }$ and $\lceil T_i^{ \eoq } \rceil^{ (\tilde{\Delta}) }$. It is not difficult to verify that, since $\hat{\Delta} \leq \tilde{\Delta} \leq (1 + \eps) \cdot \hat{\Delta}$, we are exceeding the long-run average cost of $\hat{T}$ by a factor of at most $1 + \eps$.  

%% file: TEX-Resource-Constrained.tex
\section{Resource-Constrained JRP: The General Setting} \label{sec:RC_general}

In what follows, we prove that the resource-constrained joint replenishment problem can be efficiently approximated within factor $1.417$ of optimal, thereby establishing Theorem~\ref{thm:approx_JRP_general_RC}. For this purpose, Section~\ref{subsec:overview_resource_general} focuses on presenting the high-level ideas of our approach, whose specifics are provided in Sections~\ref{subsec:procedure_slight_shift} and~\ref{subsec:procedure_rand_shift}. 

\subsection{Algorithmic overview} \label{subsec:overview_resource_general}

\paragraph{Convex relaxation.} Toward deriving Theorem~\ref{thm:approx_JRP_general_RC}, our starting point is similar to that of earlier papers \citep{Roundy89, TeoB01}. Namely, we begin by computing an optimal solution $T^* = (T_{\min}^*, T_1^*, \ldots, T_n^*)$ to the following convex relaxation:
\begin{equation} \label{eqn:conv_relax_RCJRP} \tag{RC} \begin{array}{lll}
\min & {\displaystyle \frac{ K_0 }{ T_{\min} } + \sum_{i \in [n]} \left( \frac{ K_i }{ T_i } + H_i T_i \right)} \qquad \\
\text{s.t.} & T_i \geq T_{\min} & \forall \, i \in [n] \\
& {\displaystyle \sum_{i \in [n]} \frac{ \alpha_{id} }{ T_i } \leq \beta_d} & \forall \, d \in [D]
\end{array}
\end{equation}

\paragraph{Classifying commodities.} With respect to $T^*$, we say that commodity $i \in [n]$ is small when $T_i^* \leq \frac{ 8 }{ 7 } \cdot T_{\min}^*$; otherwise, this commodity is large. The sets of such commodities will be denoted by ${\cal S}^*$ and ${\cal L}^*$, respectively, with the convention that their contributions toward the EOQ-based component of  $\opt\eqref{eqn:conv_relax_RCJRP}$ are designated by
\[ C({\cal S}^*) ~~=~~ \sum_{i \in {\cal S}^*} C_i( T_i^* ) \qquad \text{and} \qquad C({\cal L}^*) ~~=~~ \sum_{i \in {\cal L}^*} C_i( T_i^* ) \ . \]

\paragraph{Two rounding procedures.} In Section~\ref{subsec:procedure_slight_shift}, we describe a deterministic rounding procedure for computing a replenishment policy $\tilde{T}^A$, being particularly appealing when $\frac{ K_0 }{ T_{\min}^* }$ and $C({\cal S}^*)$ constitute a large chunk of $\opt\eqref{eqn:conv_relax_RCJRP}$. Specifically, the operational cost of this policy will be
\begin{equation} \label{eqn:RC_general_3}
F( \tilde{T}^A ) ~~\leq~~ \frac{ 7 }{ 8 } \cdot \frac{ K_0 }{T_{\min}^* } + \frac{ 8 }{ 7 } \cdot C( {\cal S}^* ) + 2 \cdot C( {\cal L}^* ) \ .    
\end{equation}
Then, in Section~\ref{subsec:procedure_rand_shift}, we devise a randomized rounding procedure for computing a  replenishment policy $\tilde{T}^B$, that will be useful when $C({\cal L}^*)$ forms a large fraction of $\opt\eqref{eqn:conv_relax_RCJRP}$. Here, we will show that the expected operational cost of this policy is
\begin{equation} \label{eqn:RC_general_4}
\ex{ F( \tilde{T}^B ) } ~~\leq~~ \frac{ 1 }{ \ln 2 } \cdot \frac{ K_0 }{ T_{\min}^* } + \frac{ 1 }{ \ln 2 } \cdot C( {\cal S}^* ) + \left( 1 + \frac{ 1 }{ 4\ln 2 } \right)  \cdot C( {\cal L}^* ) \ . 
\end{equation}

\paragraph{Approximation guarantee.} By picking the cheaper of these two policies, it follows that our expected cost is 
\begin{eqnarray}
&& \ex{ \min \left\{ F( \tilde{T}^A ), F( \tilde{T}^B ) \right\} } \nonumber \\
&& \qquad \leq~~ \min \left\{ F( \tilde{T}^A ), \ex{ F( \tilde{T}^B ) } \right\} \label{eqn:RC_general_1} \\
&& \qquad \leq~~ 0.087 \cdot \left( \frac{ 7 }{ 8 } \cdot \frac{ K_0 }{ T_{\min}^* } + \frac{ 8 }{ 7 } \cdot C( {\cal S}^* ) + 2 \cdot C( {\cal L}^* ) \right) \nonumber \\
&& \qquad \qquad \mbox{} + (1 - 0.087) \cdot \left( \frac{ 1 }{ \ln 2 } \cdot \frac{ K_0 }{ T_{\min}^* } + \frac{ 1 }{ \ln 2 } \cdot C( {\cal S}^* ) + \left( 1 + \frac{ 1 }{ 4\ln 2 } \right) \cdot C( {\cal L}^* ) \right) \label{eqn:RC_general_2} \\
&& \qquad \leq~~ 1.394 \cdot \frac{ K_0 }{ T_{\min}^* } + 1.417 \cdot C( {\cal S}^* ) +  1.417 \cdot C( {\cal L}^* ) \nonumber \\
&& \qquad \leq~~ 1.417 \cdot \opt\eqref{eqn:conv_relax_RCJRP} \ . \nonumber 
\end{eqnarray}
Here, inequality~\eqref{eqn:RC_general_1} follows from Jensen's inequality, noting that the function $x \mapsto \min \{ c, x \}$ is concave for any fixed $c \in \bbR$. Inequality~\eqref{eqn:RC_general_2} is obtained by substituting~\eqref{eqn:RC_general_3} and~\eqref{eqn:RC_general_4}.
     
\subsection{The right-shift procedure} \label{subsec:procedure_slight_shift}

Our first policy, $\tilde{T}^A$, is designed to take advantage of the scenario where a large fraction of the relaxed optimum $\opt\eqref{eqn:conv_relax_RCJRP}$ is coming from the joint ordering term $\frac{ K_0 }{ T_{\min}^* }$ or from the EOQ-based component $C({\cal S}^*)$ of small commodities. This policy ensures that both ingredients will incur appropriately-bounded rounding errors by right-shifting the time intervals of small commodities to the threshold point $\frac{ 8 }{ 7 } \cdot T_{\min}^*$ along the following lines:  
\begin{itemize}
    \item {\em Placing joint orders:} Joint orders will be placed at integer multiples of $\Delta = \frac{ 8 }{ 7 } \cdot T_{\min}^*$. As such, we ensure that $\tilde{T}^A$ has a long-run ordering cost of 
    \[ J( \tilde{T}^A ) ~~=~~ \frac{ K_0 }{ \Delta } ~~=~~ \frac{ 7 }{ 8 } \cdot \frac{ K_0 }{ T_{\min}^* } \ . \]

    \item {\em Placing commodity-specific orders:} For every commodity $i \in [n]$, we set its time interval $\tilde{T}^A_i$ by rounding $T_i^*$ up to the nearest integer multiple of $\Delta$, meaning that $\tilde{T}^A_i = \lceil T_i^* \rceil^{ (\Delta) }$. Consequently, since $T_i^* \leq \frac{ 8 }{ 7 } \cdot T_{\min}^* = \Delta$ for every small commodity $i \in {\cal S}^*$, we have $\tilde{T}^A_i = \Delta$, implying in turn that the EOQ-based cost of this commodity is
    \[ C_i( \tilde{T}^A_i ) ~~=~~ \frac{ K_i }{ \Delta } + H_i \Delta ~~\leq~~ \frac{ K_i }{ T_i^* } + \frac{ 8 }{ 7 } \cdot H_i T_i^* ~~\leq~~ \frac{ 8 }{ 7 } \cdot C_i( T_i^* ) \ . \]
    On the other hand, for every large commodity $i \in {\cal L}^*$, 
    \[ C_i( \tilde{T}^A_i ) ~~=~~ \frac{ K_i }{ \lceil T_i^* \rceil^{ (\Delta) } } + H_i \lceil T_i^* \rceil^{ (\Delta) } ~~\leq~~ \frac{ K_i }{ T_i^* } +  H_i \cdot ( T_i^* + \Delta ) ~~\leq~~ \frac{ K_i }{ T_i^* } +  2 H_i T_i^* ~~\leq~~ 2 \cdot C_i( T_i^* ) \ ,  \]
    where the next-to-last inequality holds since $T_i^* > \frac{ 8 }{ 7 } \cdot T_{\min}^* = \Delta$.    
\end{itemize}
By combining these observations, it follows that the long-run operational cost of our policy is
\[ F( \tilde{T}^A ) ~~\leq~~ \frac{ 7 }{ 8 } \cdot \frac{ K_0 }{ T_{\min}^* } + \frac{ 8 }{ 7 } \cdot C( {\cal S}^* ) + 2 \cdot C( {\cal L}^* ) \ ,  \]
which is precisely the bound stated in~\eqref{eqn:RC_general_3}. Moreover, it is important to emphasize that any replenishment policy $\tilde{T}$ with $\tilde{T} \geq T^*$ is necessarily resource-feasible. Indeed, for every $d \in [D]$, we have $\sum_{i \in [n]} \frac{ \alpha_{id} }{ \tilde{T}_i } \leq \sum_{i \in [n]} \frac{ \alpha_{id} }{ T_i^* } \leq \beta_d $, where the latter inequality holds since $T^*$ forms a feasible solution to~\eqref{eqn:conv_relax_RCJRP}.    

\subsection{The randomized-shift procedure} \label{subsec:procedure_rand_shift}

Here, we consider the complementary scenario, where a large fraction of $\opt\eqref{eqn:conv_relax_RCJRP}$ is coming from the EOQ-based component of large commodities, $C({\cal L}^*)$. In particular, we examine the natural idea of constructing an evenly-spaced policy, whose spacing parameter is obtained via the randomized rounding approach of \citet{TeoB01}. While this procedure may scale up the contribution of small commodities by $\frac{ 1 }{ \ln 2 } \approx 1.442$, the crux of our refined analysis would be to prove that large commodities are scaled by at most $1 + \frac{ 1 }{ 4\ln 2 } \approx 1.36$. To this end, our current policy $\tilde{T}^B$ will be defined as follows:
\begin{itemize}
    \item {\em Placing joint orders:} Joint orders will be placed at integer multiples of the random base $\Delta_U = \frac{ T_{\min}^* }{ e^U }$, where $U \sim U(0, \ln 2)$.

    \item {\em Placing commodity-specific orders:} For every commodity $i \in [n]$, we set its time interval $\tilde{T}^B_i$ by rounding $T_i^*$ up to the nearest integer multiple of $\Delta_U$, meaning that $\tilde{T}^B_i = \lceil T_i^* \rceil^{ (\Delta_U) }$.
\end{itemize}
We mention in passing that $\tilde{T}^B$ is a resource-feasible policy, for any realization of $U$. Similarly to Section~\ref{subsec:procedure_slight_shift}, this claim follows by noting that $\tilde{T}^B \geq T^*$. 

\paragraph{Analysis.} To analyze the performance guarantee of our policy, we begin by listing a number of auxiliary observations. 

\begin{claim} \label{clm:bound_round_eU}
$\exsubpar{ U }{ e^U } = \frac{ 1 }{ \ln 2 }$ and  $\exsubpar{ U }{ e^{-U} } = \frac{ 1 }{ 2\ln 2 }$. 
\end{claim}

\begin{claim} \label{clm:bound_round_1_3by2}
When $T_i^* \in [T_{\min}^*, \frac{ 3 }{ 2 } \cdot T_{\min}^*]$, we have $\exsubpar{ U }{ \frac{ \tilde{T}^B_i }{ T_i^*} } = \frac{ 1 }{ 2\ln 2} \cdot ( 1 +  \frac{ T_{\min}^* }{ T_i^* } )$.
\end{claim}

\begin{claim} \label{clm:bound_round_3by2_2}
When $T_i^* \in (\frac{ 3 }{ 2 } \cdot T_{\min}^*, 2T_{\min}^*]$, we have $\exsubpar{ U }{ \frac{ \tilde{T}^B_i }{ T_i^*} } = \frac{ 5 }{ 6\ln 2}$.
\end{claim}

Claim~\ref{clm:bound_round_eU} follows from one-line calculations, and we omit its straightforward proof. Claims~\ref{clm:bound_round_1_3by2} and~\ref{clm:bound_round_3by2_2} require finer arguments, and we provide their proofs in Appendices~\ref{app:proof_clm_bound_round_1_3by2} and~\ref{app:proof_clm_bound_round_3by2_2}, respectively. Given these observations, we are now ready to establish the upper bound~\eqref{eqn:RC_general_4} on the long-run operational cost of our policy.

\begin{lemma}
$\exsubpar{U}{ F( \tilde{T}^B ) } \leq \frac{ 1 }{ \ln 2 } \cdot \frac{ K_0 }{ T_{\min}^* } + \frac{ 1 }{ \ln 2 } \cdot C( {\cal S}^* ) + ( 1 + \frac{ 1 }{ 4\ln 2 } )  \cdot C( {\cal L}^* )$.
\end{lemma}
\begin{proof}
To derive the desired bound on the expected operational cost of $\tilde{T}^B$, we first observe that in terms of joint orders,
\[ \exsub{ U }{ J( \tilde{T}^B ) } ~~=~~ \exsub{ U }{\frac{ K_0 }{ \Delta_U } } ~~=~~ \exsub{ U }{ e^U } \cdot \frac{ K_0 }{ T_{\min}^* } ~~=~~ \frac{ 1 }{ \ln 2 } \cdot \frac{ K_0 }{ T_{\min}^* } \ , \]
where the last equality follows from Claim~\ref{clm:bound_round_eU}.

Moving on to consider EOQ-based costs, we remind the reader that $T_i^* \in [T_{\min}^*, \frac{ 8 }{ 7 } \cdot T_{\min}^*]$ for every small commodity $i \in {\cal S}^*$. Therefore, this time interval satisfies the condition of Claim~\ref{clm:bound_round_1_3by2}, implying that $\exsubpar{ U }{ \frac{ \tilde{T}^B_i }{ T_i^*} } = \frac{ 1 }{ 2\ln 2} \cdot ( 1 +  \frac{ T_{\min}^* }{ T_i^* } ) \leq \frac{ 1 }{ \ln 2}$. Consequently,
\begin{eqnarray*}
\exsub{ U }{ C_i( \tilde{T}^B_i ) } & = & \exsub{ U }{ \frac{ K_i }{ \tilde{T}^B_i } + H_i \tilde{T}^B_i } \\
& \leq & \frac{ K_i }{ T_i^* } + H_i T^*_i \cdot  \exsub{ U }{ \frac{ \tilde{T}^B_i }{ T_i^* } } \\
& \leq & \frac{ K_i }{ T_i^* } + \frac{ 1 }{  \ln 2 } \cdot H_i T_i^* \\
& \leq & \frac{ 1 }{  \ln 2 } \cdot C_i( T_i^* ) \ . 
\end{eqnarray*}

Now, focusing on a single large commodity $i \in {\cal L}^*$, let us write $T_i^* = \theta_i T_{\min}^*$ for convenience of notation. Clearly, $\theta_i \geq \frac{ 8 }{ 7 }$, and we proceed to consider the next few cases, depending on the magnitude of this parameter.
\begin{enumerate}
    \item {\em When $\theta_i \in [\frac{ 8 }{ 7 }, \frac{ 3 }{ 2 }]$:} Here, we fall into the regime of Claim~\ref{clm:bound_round_1_3by2}, and thus $\exsubpar{ U }{ \frac{ \tilde{T}^B_i }{ T_i^*} } = \frac{ 1 }{ 2\ln 2} \cdot ( 1 +  \frac{ 1 }{ \theta_i } ) \leq \frac{ 15 }{ 16 \ln 2}$. Consequently,
    \[ \exsub{ U }{ C_i( \tilde{T}^B_i ) }  ~~\leq~~ \frac{ K_i }{ T_i^* } + H_i T^*_i \cdot  \exsub{ U }{ \frac{ \tilde{T}^B_i }{ T_i^* } } ~~\leq~~ \frac{ K_i }{ T_i^* } + \frac{ 15 }{ 16 \ln 2} \cdot H_i T_i^* ~~\leq~~ \frac{ 15 }{ 16 \ln 2} \cdot C_i( T_i^* ) \ . \] 

    \item {\em When $\theta_i \in (\frac{ 3 }{ 2 },2]$:} In this case, we are within the regime of Claim~\ref{clm:bound_round_3by2_2}, implying that $\exsubpar{ U }{ \frac{ \tilde{T}^B_i }{ T_i^*} } = \frac{ 5 }{ 6\ln 2}$. Therefore, following the same logic as in item~1 above, $\exsubpar{ U }{ C_i( \tilde{T}^B_i ) }  \leq \frac{ 5 }{ 6 \ln 2 } \cdot C_i( T_i^* )$. 

     \item{\em When $\theta_i > 2$:} Here, since $\tilde{T}^B_i = \lceil T_i^* \rceil^{ (\Delta_U) }$, we have
     \[ \exsub{ U }{ \tilde{T}^B_i } ~~\leq~~ \exsub{ U }{ T_i^* + \Delta_U } ~~=~~ T_i^* + T_{\min}^* \cdot \exsub{ U }{ e^{-U} } ~~<~~ \left( 1 + \frac{ 1 }{ 4\ln 2 }  \right) \cdot T_i^* \ , \]
     where the second inequality is obtained by combining Claim~\ref{clm:bound_round_eU} with our case hypothesis of $\theta_i > 2$. Consequently,
     \[ \exsub{ U }{ C_i( \tilde{T}^B_i ) }  ~~\leq~~ \frac{ K_i }{ T_i^* } + \left( 1 + \frac{ 1 }{ 4\ln 2 }  \right) \cdot H_i T_i^* ~~\leq~~ \left( 1 + \frac{ 1 }{ 4\ln 2 }  \right) \cdot C_i( T_i^* ) \ . \] 
\end{enumerate}
In summary, we infer that the expected  EOQ-based cost of our policy is
\begin{eqnarray*}
\exsub{ U }{ \sum_{i \in [n]} C_i( \tilde{T}^B_i ) } & \leq & \frac{ 1 }{ \ln 2 } \cdot C( {\cal S}^* ) + \max \left\{ \frac{ 15 }{ 16 \ln 2}, \frac{ 5 }{ 6 \ln 2 } , 1 + \frac{ 1 }{ 4\ln 2 } \right\}  \cdot C( {\cal L}^* ) \\
& = & \frac{ 1 }{ \ln 2 } \cdot C( {\cal S}^* ) + \left( 1 + \frac{ 1 }{ 4\ln 2 } \right)  \cdot C( {\cal L}^* ) \ .   
\end{eqnarray*}    
\end{proof}

%%%%%%%%%%%%%%%%%%%%%%%%%%%%%%%%%
\section{Resource-Constrained JRP: \bstitle{O(1)} Constraints} \label{sec:RC_constant}

In this section, we look into the type of performance guarantees that can be attained, when running times are allowed to be exponential in the number of resource constraints, $D$. Theorem~\ref{thm:approx_JRP_O1_RC} summarizes our main result in this direction, arguing that the resource-constrained joint replenishment problem can be approximated within factor $1 + \eps$ in $O( n^{\tilde{O}( D^3/\eps^4 )} )$ time. Toward this objective, Section~\ref{subsec:RC_discretize} explains how one could go about discretizing the seemingly-continuous decision space of replenishment policies, thereby arriving at an integer programming formulation. Section~\ref{subsec:LP_relax_O1_constraints} proposes an enumeration-based  method to strengthened the resulting linear relaxation. Finally, Section~\ref{subsec:RC_O1constraints_round} presents our randomized rounding procedure, showing that with constant probability, we indeed compute a near-optimal policy.

\subsection{Discretization} \label{subsec:RC_discretize}

In what follows, we sketch the main ideas behind our construction of efficient discretization sets for the resource-constrained joint replenishment problem. In particular, rather than allowing the time intervals $T_1, \ldots, T_n$ of a given policy to take arbitrary non-negative values, we convert our decision to be combinatorial, by restricting these intervals to a respective collection of finite sets, ${\cal T}_1, \ldots, {\cal T}_n$.

\paragraph{Construction.} In the absence of resource constraints, our method for placing commodity-specific orders within $\Psi$-pairwise alignment (see Section~\ref{subsec:construct_policy}) can retrospectively be viewed as creating a discrete set ${\cal T}_i$ for each commodity $i$. The latter set consists of the approximate representatives $\{ \tilde{R}_{\ell} \}_{ \ell \in {\cal A}^* }$, along with a single ``large'' interval, $\lceil T^{\max}_i \rceil^{ (\tilde{R}_1) }$, where $T^{\max}_i = \max \{ \frac{ 1 }{ \eps } \cdot \tilde{T}_{\min}, \sqrt{ K_i / H_i  } \}$. 
Similarly, when resource constraints are present, the  representatives $\{ \tilde{R}_{\ell} \}_{ \ell \in {\cal A}^* }$ can be defined precisely as in Section~\ref{subsec:construct_policy}. However, adding only $\lceil T^{\max}_i \rceil^{ (\tilde{R}_1) }$ will not be sufficient for our purposes, and we therefore augment ${\cal T}_i$ with $O( \frac{ 1 }{ \eps } \log \frac{ n }{ \eps } )$ extra options. 

Specifically, let us first observe that each constraint $\sum_{i \in [n]} \frac{ \alpha_{id} }{ T_i } \leq \beta_d$ implies in particular that $T_i \geq \frac{ \alpha_{id} }{ \beta_d }$. Consequently, any resource-feasible policy is operating under a lower bound of the form $T_i \geq T^{\myLB}_i = \max_{d \in [D]} \frac{ \alpha_{id} }{ \beta_d }$. Yet another important remark is that, whenever we pick $T_i \geq \frac{ n }{ \eps } \cdot T^{\myLB}_i$, this commodity has a tiny contribution toward each constraint, in the sense that $\frac{ \alpha_{id} }{ T_i } \leq \frac{ \eps }{ n } \cdot \beta_d$. Motivated by these observations, let $T_i^{(1)}, \ldots, T_i^{(P)}$ be the sequence of points that partition the segment $[T^{\myLB}_i, \frac{ n }{ \eps } \cdot T^{\myLB}_i ]$ by powers of $1 + \eps$. Namely,    
\[ T_i^{(1)} ~~=~~ T^{\myLB}_i, \quad  T_i^{(2)} ~~=~~ (1 + \eps) \cdot T^{\myLB}_i, \quad \ldots \]
so on and so forth, where in general $T_i^{(p)} = (1 + \eps)^{p-1} \cdot T^{\myLB}_i$. Here, $P$ is the minimal integer $p$ for which $(1 + \eps)^{p-1} \geq \frac{ n }{ \eps }$, meaning that $P = O( \frac{ 1 }{ \eps } \log \frac{ n }{ \eps } )$. Now, for every $p \in [P]$ with $T_i^{(p)} \geq \frac{ 1 }{ \eps } \cdot \tilde{T}_{\min}$, we augment ${\cal T}_i$ with the time interval $\lceil T_i^{(p)} \rceil^{ (\tilde{R}_1) }$. 

\paragraph{Guaranteed properties.} We proceed by listing the main structural properties of the discretization sets ${\cal T}_1, \ldots, {\cal T}_n$, which will be instrumental in subsequent sections. 
\begin{enumerate}
    \item {\em Logarithmic size:} By the preceding discussion, we know that each set ${\cal T}_i$ consists of at most $| {\cal A}^* | + P + 1$ time intervals, meaning that $| {\cal T}_i | = O( \frac{ 1 }{ \eps } \log \frac{ n }{ \eps } )$, for every commodity $i \in [n]$.

    \item {\em Joint orders are not an issue:} It is important to emphasize that each of the values $\lceil T^{\max}_i \rceil^{ (\tilde{R}_1) }, \lceil T_i^{(1)} \rceil^{ (\tilde{R}_1) }, \ldots, \lceil T_i^{(P)} \rceil^{ (\tilde{R}_1)}$ falls on an integer multiple of $\tilde{R}_1$, meaning that we cannot be creating joint order beyond those of the approximate representatives $\{ \tilde{R}_{\ell} \}_{ \ell \in {\cal A}^* }$. As a result, any policy $T = (T_1, \ldots, T_n) \in {\cal T}_1 \times \cdots \times {\cal T}_n$ has a long-run joint ordering cost of $J( T ) \leq J( \tilde{\cal R} ) \leq (1 + \eps) \cdot J( T^* )$.

    \item {\em Good EOQ-costs are achievable:} Our construction  ensures that there exists a $(1 + \eps)$-feasible policy $T = (T_1, \ldots, T_n) \in {\cal T}_1 \times \cdots \times {\cal T}_n$ whose combined EOQ-based cost is $\sum_{i \in [n]} C_i( T_i ) \leq (1 + \eps) \cdot \sum_{i \in [n]} C_i( T_i^* )$. Here, $(1 + \eps)$-feasibility means that each resource constraint is exceeded by a factor of at most $1 + \eps$, i.e., $\sum_{i \in [n]} \frac{ \alpha_{id} }{ T_i } \leq (1 + \eps) \cdot \beta_d$. The arguments in this context have a flavor very similar to those presented in Section~\ref{subsec:cost_comm_orders}, and are therefore not repeated here. In essence, when $T_i^* \leq  \frac{ 1 }{ \eps } \cdot \tilde{T}_{\min}$, this time interval will be replaced by the appropriate representative in $\{ \tilde{R}_{\ell} \}_{ \ell \in {\cal A}^* }$. When $T_i^* \in (\frac{ 1 }{ \eps } \cdot \tilde{T}_{\min}, \frac{ n }{ \eps } \cdot T^{\myLB}]$, its replacement will be the nearest $\lceil T_i^{(p)} \rceil^{ (\tilde{R}_1) }$. Finally, when $T_i^* > \frac{ n }{ \eps } \cdot T^{\myLB}$, one of $\lceil T_i^{(P)} \rceil^{ (\tilde{R}_1) }$ and $\lceil T^{\max}_i \rceil^{ (\tilde{R}_1) }$ is the right choice. 
\end{enumerate}

\paragraph{IP formulation.} Given property~2 above,  we will not be worried about long-run joint ordering costs. Instead, our attention will be focused on picking the time intervals $T_1, \ldots, T_n$ out of ${\cal T}_1, \ldots, {\cal T}_n$, respectively, with the objective of minimizing long-run EOQ-based costs. Moving forward, it is  convenient to formulate the latter question as an integer program. To this end, for every commodity $i \in [n]$ and time interval $t \in {\cal T}_i$, let $x_{it}$ be a binary decision variable, indicating whether we set $T_i = t$. As such, our problem of interest can be written as:
\begin{equation} \label{eqn:IP_RCJRP} \tag{IP} \begin{array}{lll}
\min & {\displaystyle \sum_{i \in [n]} \sum_{t \in {\cal T}_i} \left( \frac{ K_i }{ t } + H_i t \right) x_{it} }\\
\text{s.t.} & {\displaystyle \sum_{t \in {\cal T}_i} x_{it} = 1} & \forall \, i \in [n] \\
& {\displaystyle \sum_{i \in [n]} \alpha_{id} \cdot \sum_{t \in {\cal T}_i} \frac{ x_{it} }{ t } \leq (1 + \eps) \cdot \beta_d} \qquad & \forall \, d \in [D] \\
& x_{it} \in \{ 0, 1 \} & \forall \, i \in [n], t \in {\cal T}_i
\end{array}
\end{equation}
It is worth pointing out that rather than keeping each resource constraint in its original form (i.e., with $\leq \beta_d$), the program above allows for $(1 + \eps)$-feasibility. The fundamental reason behind this feature is that property~3 argues about the existence of an inexpensive $(1 + \eps)$-feasible policy. For all we know, restricting ourselves to truly feasible policies while concurrently picking from ${\cal T}_1, \ldots, {\cal T}_n$ could be significantly more expensive. Of course, we will eventually have to compute a truly feasible policy, but for the time being, we state the next result regarding the optimal value of~\eqref{eqn:IP_RCJRP}. 

\begin{observation}
$\opt\eqref{eqn:IP_RCJRP} \leq (1 + \eps) \cdot \sum_{i \in [n]} C_i( T_i^* )$.    
\end{observation}

\subsection{Linear relaxation} \label{subsec:LP_relax_O1_constraints}

Unfortunately, it is not difficult to verify that, by replacing the integrality requirement $x_{it} \in \{ 0, 1 \}$ with non-negativity constraints and nothing more, we could create an unbounded integrality gap. To address this issue, we proceed by generating two families of valid constraints, inspired by well-known ideas related to approximation schemes for the generalized assignment problem (see, for instance, \cite{JansenP01, JansenM19, KonesL19}).

\paragraph{Heavy pairs.} Letting $\delta = \frac{ \eps^4 }{ D^2 }$, for every commodity $i \in [n]$, time interval $t \in {\cal T}_i$, and constraint $d \in [D]$, we say that $(i,t)$ is a $d$-heavy pair when $\frac{ \alpha_{id} }{ t } > \delta \beta_d$; we use ${\cal H}_d$ to denote the collection of $d$-heavy pairs. Focusing on an optimal solution $x^*$ to the discrete problem~\eqref{eqn:IP_RCJRP}, let ${\cal H}^*_d$ be the set of $d$-heavy pairs chosen by $x^*$, i.e., ${\cal H}^*_d = \{ (i,t) \in {\cal H}_d : x_{it}^* = 1 \}$. It is easy to verify that ${\cal H}^*_d < \frac{ 1+\eps }{ \delta } \leq \frac{ 2 }{ \delta }$.

\paragraph{Expensive pairs.} Similarly, for every commodity $i \in [n]$ and time interval $t \in {\cal T}_i$, we say that $(i,t)$ is an expensive pair when $\frac{ K_i }{ t } + H_i t > \eps^4  \opt\eqref{eqn:IP_RCJRP}$. We use ${\cal E}$ to denote the collection of expensive pairs, whereas ${\cal E}^*$ will designate those chosen by $x^*$, i.e., ${\cal E}^* = \{ (i,t) \in {\cal E} : x_{it}^* = 1 \}$. Once again, it is easy to verify that ${\cal E}^* < \frac{ 1 }{ \eps^4 }$. 

\paragraph{Guessing.} We proceed by guessing the identity of ${\cal H}^*_d$, for every $d \in [D]$, noting that there are $O( (n \cdot \max_{i \in [n]} |{\cal T}_i|)^{O(D/\delta)} )$ options to enumerate over. In addition, we similarly guess all members of the set ${\cal E}^*$, for which there are $O( (n \cdot \max_{i \in [n]} |{\cal T}_i|)^{O(1/\eps^4)} )$ possible options to consider. As mentioned in Section~\ref{subsec:RC_discretize}, we have $| {\cal T}_i | = O( \frac{ 1 }{ \eps } \log \frac{ n }{ \eps } )$ for every commodity $i \in [n]$, meaning that our overall number of guesses is
\[ \left( n \cdot \max_{i \in [n]} |{\cal T}_i| \right)^{O( \max \{ D / \delta , 1 / \eps^4 \})} ~~=~~ \left( \frac{ n }{ \eps } \log \left( \frac{ n }{ \eps } \right) \right)^{O( D^3 / \eps^4 )} ~~=~~ n^{\tilde{O}( D^3 / \eps^4 )} \ . \]

\paragraph{The linear relaxation.} Given these guesses, we can infer two families of valid constraints with respect to~\eqref{eqn:IP_RCJRP}. First, for every $d \in [D]$, the collection of $d$-heavy pairs to be selected is precisely ${\cal H}^*_d$, namely, $x_{it} = 1$ for $(i,t) \in {\cal H}^*_d$, and concurrently $x_{it} = 0$ for  $(i,t) \in {\cal H}_d \setminus {\cal H}^*_d$. Second, the collection of expensive pairs to be selected is ${\cal E}^*$, implying that $x_{it} = 1$ for $(i,t) \in {\cal E}^*$, whereas $x_{it} = 0$ for $(i,t) \in {\cal E} \setminus {\cal E}^*$. By plugging these equations back into~\eqref{eqn:IP_RCJRP} and replacing our integrality requirement $x_{it} \in \{ 0, 1 \}$ with a non-negativity constraint, $x_{it} \geq 0$, we obtain the following linear relaxation:
\begin{equation} \label{eqn:linear_relax_RCJRP} \tag{LP} \begin{array}{lll}
\min & {\displaystyle \sum_{i \in [n]} \sum_{t \in {\cal T}_i} \left( \frac{ K_i }{ t } + H_i t \right) x_{it} }\\
\text{s.t.} & {\displaystyle \sum_{t \in {\cal T}_i} x_{it} = 1} & \forall \, i \in [n] \\
& {\displaystyle \sum_{i \in [n]} \alpha_{id} \cdot \sum_{t \in {\cal T}_i} \frac{ x_{it} }{ t } \leq (1 + \eps) \cdot \beta_d} \qquad & \forall \, d \in [D] \\
& x_{it} = 1 & \forall \, d \in [D], (i,t) \in {\cal H}^*_d \\
& x_{it} = 0 & \forall \, d \in [D], (i,t) \in {\cal H}_d \setminus {\cal H}^*_d \\
& x_{it} = 1 & \forall \, (i,t) \in {\cal E}^* \\
& x_{it} = 0 & \forall \, (i,t) \in {\cal E} \setminus {\cal E}^* \\
& x_{it} \geq 0 & \forall \, i \in [n], t \in {\cal T}_i
\end{array}
\end{equation}

\subsection{The randomized rounding procedure} \label{subsec:RC_O1constraints_round}

Given an optimal solution $\tilde{x}$ to the linear relaxation~\eqref{eqn:linear_relax_RCJRP}, we construct a random replenishment policy $\tilde{T} = (\tilde{T}_1, \ldots, \tilde{T}_n)$ as follows. For every commodity $i \in [n]$, due to having $\sum_{t \in {\cal T}_i} x_{it} = 1$ as a constraint of~\eqref{eqn:linear_relax_RCJRP}, we can view $( \tilde{x}_{it} )_{t \in {\cal T}_i}$ as probabilities. As such, we choose a random time interval $\tilde{T}_i$ out of ${\cal T}_i$ according to these probabilities. It is important to emphasize that $\tilde{T}_1, \ldots, \tilde{T}_n$ are independently drawn. 

\paragraph{Cost analysis.} We first argue that, with constant probability, our  policy $\tilde{T}$ has a near-optimal cost. The exact statement of this claim is laid down by Lemma~\ref{lem:UB_prob_large_cost} below, whose proof is presented in Appendix~\ref{app:proof_lem_UB_prob_large_cost}. The general intuition behind our proof is that, since $\tilde{T}$ is  forced to select the collection ${\cal E}^*$ of expensive pairs and to avoid selecting any other expensive pair, meaningful cost deviations can only be attributed to inexpensive pairs. However, since any such pair $(i,t)$ has $\frac{ K_i }{ t } + H_i t \leq \eps^4  \opt\eqref{eqn:IP_RCJRP}$, an appropriate concentration inequality will show that an $\Omega( \eps \opt\eqref{eqn:linear_relax_RCJRP} )$ deviation occurs with low probability.

\begin{lemma} \label{lem:UB_prob_large_cost}
$\prpar{ \sum_{i \in [n]} ( \frac{ K_i }{ \tilde{T}_i } + H_i \tilde{T}_i ) \leq (1 + \eps) \cdot \opt\eqref{eqn:linear_relax_RCJRP} } \geq \frac{ 2 }{ 3 }$. 
\end{lemma}

\paragraph{Feasibility analysis.} Our next claim is that, with constant probability, the policy $\tilde{T}$ is nearly feasible. To formalize this claim, we first derive an upper bound on the probability of violating any given resource constraint by a factor greater than $1 + 3\eps$. The latter bound can be established similarly to how we prove Lemma~\ref{lem:UB_prob_large_cost}; for completeness, its finer details are provided in Appendix~\ref{app:proof_lem_UB_prob_large_resource}.   

\begin{lemma} \label{lem:UB_prob_large_resource} 
$\prpar{ \sum_{i \in [n]} \frac{ \alpha_{id} }{ \tilde{T}_i } \geq (1+3\eps) \cdot \beta_d} \leq \frac{ 1 }{ 3D }$, for every $d \in D$. 
\end{lemma}

Now, by taking the union bound over all $d \in [D]$, we conclude that $\tilde{T}$ is a $(1+3\eps)$-feasible policy with probability at least $2/3$. Combined with Lemma~\ref{lem:UB_prob_large_cost}, it follows that this policy is $(1 + \eps)$-approximate at the same time, with probability at least $1/3$. Finally, to ensure true resource-feasibility, we simply scale $\tilde{T}$ up by a factor of $1 +3\eps$, blowing its EOQ-based cost by at most $1 +3\eps$ as well.

%% file: TEX-Remarks.tex
\section{Concluding Remarks} \label{sec:conclusions}

We believe that our work introduces a wide array of topics to be investigated in future research. The next few paragraphs are dedicated to highlighting some of these prospective directions, ranging from seemingly doable to highly non-trivial. 

\paragraph{Fine-grained approximations?} As mentioned in Section~\ref{subsec:contributions}, to communicate our main ideas in a digestible way, we have not made concentrated efforts to push the envelope of constant-factor approximations. In essence, the performance guarantees stated in Theorems~\ref{thm:fixed_base_main}, \ref{thm:even_spaced_better}, and~\ref{thm:approx_JRP_general_RC} should be viewed more of as proof-of-concepts rather than as genuine attempts to arrive at the best-possible approximations. As part of future research, it would be interesting to examine whether highly refined analysis, possibly combined with additional ideas, could perhaps lead to improved guarantees for the fixed-base joint replenishment problem, for the performance of evenly-spaced policies, and for resource-constrained models. At present time, we only know of subtle improvements along these lines, all coming at the cost of lengthy and involved arguments. 

\paragraph{The fixed-base setting: Approximation scheme?} As expert readers have surely observed, Theorem~\ref{thm:fixed_base_main} allows us to readily migrate performance guarantees for the variable-base model to its fixed-base counterpart, conditional on going through the convex relaxation route \citep{Roundy85, Roundy86, JacksonMM85}. That said, our previous work in this context \citep{Segev23JRP}, along with Section~\ref{sec:EPTAS-variable-base} of the current paper, reopens this gap via an approximation scheme for the variable-base setting versus a $(\frac{ 1 }{ \sqrt{2} \ln 2 } + \eps)$-approximation for the fixed-base setting. Toward further progress, one fundamental question is whether the algorithmic ideas behind $\Psi$-pairwise alignment can be leveraged to deal with the fixed-based restriction. Among other technical issues, it is still unclear how one should go about efficiently defining the approximate representatives $\tilde{\cal R} = \{ \tilde{R}_{\ell} \}_{ \ell \in {\cal A}^* }$, as in step~5 of Section~\ref{subsec:construct_policy}, while limiting their values to integer multiples of the prespecified base $\Delta$.

\paragraph{Additional long-standing open questions.} Given our improved understanding of various joint replenishment models, a particularly challenging direction for future research would be to make analogous progress with respect to additional lot sizing problems. From this perspective, two cornerstone problems that appear to be natural candidates are:
\begin{itemize}
\item Dynamic replenishment policies for single-resource multi-item inventory systems.

\item Staggering policies for multi-cycle multi-item inventory systems. 
\end{itemize}
Even though both settings have attracted a great deal of attention for decades, in terms of provably-good performance guarantees, our progress has plateaued upon landing at $2$-approximate SOSI-policies, emerging from the ingenious work of \cite{Anily91} and \cite{GallegoQS96}. To learn more about the rich history of these problems and about some of their analytical obstacles, we refer avid readers to the book chapter of \citet[Sec.~9.2]{SimchiLeviCB} on this topic.

%% file: TEX-More-Proofs.tex
\section{Additional Proofs from Sections~\ref{sec:intro}-\ref{sec:EPTAS-variable-base}} 

\subsection{Proof of Claim~\ref{clm:EOQ_properties}(4)} \label{app:proof_clm_EOQ_properties_4}

Let us first observe that $C( \alpha ) = \frac{ T }{\alpha} \cdot \frac{ K }{ T } + \frac{\alpha}{ T } \cdot HT$ and $C( \beta ) = \frac{ T }{\beta} \cdot \frac{ K }{ T } + \frac{\beta}{ T } \cdot HT$. Consequently, for every $\theta \in [0,1]$, we have
\[ \min \{ C( \alpha ), C( \beta  ) \} ~~\leq~~ \left( \theta \cdot \frac{ T }{\alpha} + (1 - \theta) \cdot \frac{ T }{\beta} \right) \cdot \frac{ K }{ T } + \left( \theta \cdot \frac{\alpha}{ T } + (1 - \theta) \cdot \frac{\beta}{ T } \right) \cdot HT \ . \] 
By plugging in $\theta = \frac{ \frac{ \beta }{ T } - \frac{ T }{ \beta } }{ \frac{ \beta }{ T } - \frac{ T }{ \beta } + \frac{ T }{ \alpha } - \frac{ \alpha }{ T } } \in [0,1]$ and simplifying the bound attained, it follows that 
\[ \min \{ C( \alpha ), C( \beta  ) \} ~~\leq~~ \frac{ \alpha + \beta }{ \frac{ \alpha \beta }{ T } + T} \cdot C(T) ~~\leq~~ \frac{ 1 }{ 2 } \cdot \left( \sqrt{ \frac{ \beta }{ \alpha } } + \sqrt{ \frac{ \alpha }{ \beta } } \right) \cdot C(T) \ . \]
To verify the last inequality, one can show by elementary calculus that $\max \{ \frac{ \alpha \beta }{ \frac{ \alpha \beta }{ T } + T} : T \in [\alpha,\beta] \}$ is attained at $T^* = \sqrt{\alpha \beta}$.

\subsection{Proof of Claim~\ref{clm:Exact_opt_prob_crossing}} \label{app:proof_clm_Exact_opt_prob_crossing}

To show that $\opt\eqref{eqn:opt_prob_crossing} = \frac{ \Lambda - 1 }{ 2 \Lambda } \cdot |{\cal A}^*|^2$, it suffices to argue that we obtain an optimal solution to~\eqref{eqn:opt_prob_crossing} by uniformly setting $x_{\lambda}^* = \frac{ |{\cal A}^*| }{ \Lambda }$ for every $\lambda \in [\Lambda]$, since 
\[ \sum_{ \MyAbove{ \lambda_1, \lambda_2 \in [\Lambda]: }{ \lambda_1 < \lambda_2 } } x^*_{\lambda_1} x^*_{\lambda_2} ~~=~~ \binom{ \Lambda }{ 2 } \cdot \frac{ |{\cal A}^*|^2 }{ \Lambda^2 } ~~=~~ \frac{ \Lambda - 1 }{ 2 \Lambda } \cdot |{\cal A}^*|^2 \ . \]
For this purpose, let $x^*$ be an optimal solution to~\eqref{eqn:opt_prob_crossing}, and suppose that not all coordinates of $x^*$ are $\frac{ |{\cal A}^*| }{ \Lambda }$-valued. In this case, since $\| x^* \|_1 = |{\cal A}^*|$, there exists at least one pair of coordinates, say $1$ and $2$, such that $x^*_1 < \frac{ |{\cal A}^*| }{ \Lambda }$ and $x^*_2 > \frac{ |{\cal A}^*| }{ \Lambda }$. Letting $\delta = \min \{ x^*_2 - \frac{ |{\cal A}^*| }{ \Lambda }, \frac{ |{\cal A}^*| }{ \Lambda } - x^*_1 \} > 0$, we construct a new vector $\hat{x} \in \bbR^{ \Lambda }_+$, where $\hat{x}_{1} = x_1^* + \delta$, $\hat{x}_{2} = x_2^* - \delta$, and $\hat{x}_{\lambda} = x^*_{\lambda}$ for all $\lambda \in \Lambda \setminus \{ 1,2 \}$. It is easy to verify that $\hat{x}$ is also a feasible solution to~\eqref{eqn:opt_prob_crossing}. However, we have just arrived at a contradiction to the optimality of $x^*$, since
\[ \sum_{ \MyAbove{ \lambda_1, \lambda_2 \in [\Lambda]: }{ \lambda_1 < \lambda_2 } } \hat{x}_{\lambda_1} \hat{x}_{\lambda_2} ~~=~~ \sum_{ \MyAbove{ \lambda_1, \lambda_2 \in [\Lambda]: }{ \lambda_1 < \lambda_2 } } x^*_{\lambda_1} x^*_{\lambda_2} + \delta \cdot (x_2^* - x_1^* - \delta) ~~>~~ \sum_{ \MyAbove{ \lambda_1, \lambda_2 \in [\Lambda]: }{ \lambda_1 < \lambda_2 } } x^*_{\lambda_1} x^*_{\lambda_2} \ , \]
where the latter inequality holds since $x_2^* - x_1^* \geq 2 \delta$ and $\delta > 0$.

\subsection{Proof of Lemma~\ref{lem:eval_lim_U_Delta}} \label{app:proof_lem_eval_lim_U_Delta}

According to Lemma~\ref{lem:limit_joint_order}, we have $\lim_{ \Delta \to \infty } \frac{ N( \tilde{\cal R}^{(\lambda)}, \Delta) }{ \Delta } = \sum_{{\cal N} \subseteq \tilde{\cal C}_{\lambda}} \frac{ (-1)^{ |{\cal N}| + 1 } }{ M_{\cal N} }$, where $M_{\cal N}$ designates the least common multiple of $\{ \tilde{\cal R}_{ \ell } \}_{\ell \in {\cal N}}$. Noting that the latter summation consists of $2^{ | \tilde{\cal C}_{\lambda} | } = O( 2^{ \tilde{O}(1/\eps) } )$ terms, we proceed by explaining how to evaluate each common multiple $M_{\cal N}$ in $O( 2^{ \tilde{O}(1/\eps) } )$ time. For this purpose, let us circle back to step~5 of our construction (see Section~\ref{subsec:construct_policy}). Focusing on the tree $\tilde{\cal T}_{ \lambda }$ that spans the connected component $\tilde{\cal C}_{\lambda}$, we remind the reader that $\sigma_{ \lambda }$ stands for the source vertex of this tree. We argue that the least common multiple of $\{ \tilde{\cal R}_{ \ell } \}_{\ell \in \tilde{\cal C}_{\lambda}}$ is one of $\tilde{R}_{\sigma_{ \lambda }}, 2 \cdot \tilde{R}_{\sigma_{ \lambda }}, \ldots, \Psi^{ | \tilde{\cal C}_{\lambda} |-1 } \cdot \tilde{R}_{\sigma_{ \lambda }}$, implying in particular that for any subset ${\cal N} \subseteq \tilde{\cal C}_{\lambda}$, we can determine its corresponding multiple $M_{\cal N}$ by testing each of these values, which would require only  $O( \Psi^{ | \tilde{\cal C}_{\lambda} | } ) = O( 2^{ \tilde{O}(1/\eps) } )$ time.

To better understand how the least common multiple of $\{ \tilde{\cal R}_{ \ell } \}_{\ell \in \tilde{\cal C}_{\lambda}}$ is structured, let $u_1, \ldots, u_{| \tilde{\cal C}_{\lambda} |}$ be a permutation of the vertices in $\tilde{\cal T}_{\lambda}$, obtained by rooting this tree at $u_1 = \sigma_{ \lambda }$ and then listing vertices in weakly-increasing order of their depth. We prove by induction that, for every $k \leq | \tilde{\cal C}_{\lambda} |$, the least common multiple of $\{ \tilde{\cal R}_{ \ell } \}_{\ell \in [k]}$ is one of $\tilde{R}_{\sigma_{ \lambda }}, 2 \cdot \tilde{R}_{\sigma_{ \lambda }}, \ldots, \Psi^{ k-1 } \cdot \tilde{R}_{\sigma_{ \lambda }}$. The base case of $k=1$ is trivial, since $u_1 = \sigma_{ \lambda }$. For the general case of $k \geq 2$, our induction hypothesis states that the least common multiple of $\{ \tilde{\cal R}_{ \ell } \}_{\ell \in [k-1]}$ is one of $\tilde{R}_{\sigma_{ \lambda }}, 2 \cdot \tilde{R}_{\sigma_{ \lambda }}, \ldots, \Psi^{ k-2 } \cdot \tilde{R}_{\sigma_{ \lambda }}$. The important observation is that, by definition of the permutation $u_1, \ldots, u_{| \tilde{\cal C}_{\lambda} |}$, we know that $u_k$ is connected by a tree edge to some vertex $u_{\ell}$, with $\ell < k$. As such, the approximate representatives of these vertices are $\Psi$-aligned, namely, $\alpha_{ \{ u_k, u_{\ell} \}, u_k } \cdot \tilde{R}_{u_k} = \alpha_{ \{ u_k, u_{\ell} \}, u_{\ell} } \cdot \tilde{R}_{u_{\ell}}$, and therefore, the least common multiple of $\tilde{R}_{u_k}$ and $\tilde{R}_{u_{\ell}}$ is upper-bounded by $\alpha_{ \{ u_k, u_{\ell} \}, u_{\ell} } \cdot \tilde{R}_{u_{\ell}} \leq \Psi \cdot \tilde{R}_{u_{\ell}}$. Combined with the induction hypothesis, it follows that the least common multiple of $\{ \tilde{\cal R}_{ \ell } \}_{\ell \in [k]}$ is one of $\tilde{R}_{\sigma_{ \lambda }}, 2 \cdot \tilde{R}_{\sigma_{ \lambda }}, \ldots, \Psi^{ k-1 } \cdot \tilde{R}_{\sigma_{ \lambda }}$. 

\subsection{Proof of Lemma~\ref{lem:UB_marginal_hatT} (continued)} \label{app:proof_lem_UB_marginal_hatT}

\paragraph{The medium regime: $\bs{T_i^{ \eoq } \in [T_{\min}^*, \frac{ 1 }{ \eps } \cdot T_{\min}^*]}$.} Here, the important observation is that we must have $T_i^* \in [T_{\min}^*, \frac{ 1 }{ \eps } \cdot T_{\min}^*]$ as well. To this end, while $T_i^* \geq T_{\min}^*$ is obvious, suppose on the contrary that $T_i^* > \frac{ 1 }{ \eps } \cdot T_{\min}^*$. Our claim is that the latter inequality leads to contradicting the optimality of $T^*$. Indeed, one possible way of altering $T^*$ is simply to reset the time interval of commodity $i$, replacing $T_i^*$ by $\frac{ 1 }{ \eps } \cdot T_{\min}^*$. Since this  term is an integer multiple of $T_{\min}^*$, we are clearly preserving the long-run joint ordering cost of $T^*$. However, in terms of EOQ-based cost, we  have $C_i( \frac{ 1 }{ \eps } \cdot T_{\min}^* ) < C_i( T_i^* )$. This inequality follows from an argument similar to that of the low regime, noting that $C_i$ is a strictly convex function with 
a unique minimizer at $T_i^{ \eoq }$, and that $T_i^{ \eoq } \leq \frac{ 1 }{ \eps } \cdot T_{\min}^* < T_i^*$.

Now, given that the segments $S_1^*, \ldots, S_L^*$ form a partition of $[T_{\min}^*, \frac{ 1 }{ \eps } \cdot T_{\min}^*]$, there exists an index $\ell \in [L]$ for which $T_i^* \in S_{\ell}^*$, meaning in particular that this segment is active. In turn, it follows that the approximate set $\tilde{\cal R}$ includes a representative of this segment, $\tilde{R}_{\ell}$. Once again, letting ${\cal C}_{\lambda}^*$ be the connected component of $G_{\Psi}^*$ where segment $S_{\ell}^*$ resides, we know that there exists a coefficient $\gamma_{\lambda} \in 1 \pm \eps$ such that $\tilde{R}_{\ell} = \gamma_{\lambda} \cdot R^*_{\ell}$. As explained in step~6 of  Section~\ref{subsec:construct_policy}, $\tilde{R}_{\ell}$ is one of the options considered for our time interval $\tilde{T}_i$, and due to picking the option that minimizes $C_i(\cdot)$, we have
\begin{eqnarray}
C_i( \tilde{T}_i ) & \leq & C_i( \tilde{R}_{\ell} ) \nonumber \\
& \leq & (1 + 2\eps) \cdot C_i( R^*_{\ell} ) \nonumber \\
& \leq & (1 + 2\eps) \cdot (1 + \eps) \cdot C_i( T^*_i ) \label{eqn:lem_UB_marginal_hatT_4}  \\
& \leq & (1 + 5\eps) \cdot C_i( T^*_i )  \ , \nonumber
\end{eqnarray}
where inequality~\eqref{eqn:lem_UB_marginal_hatT_4} is obtained by recalling that both $T_i^*$ and $R_{\ell}^*$ reside within the segment $S_{\ell}^*$, implying that they differ by a factor of at most $1 + \eps$.

\paragraph{The high regime $\bs{T_i^{ \eoq } > \frac{ 1 }{ \eps } \cdot T_{\min}^*}$.} As explained in step~6 of  Section~\ref{subsec:construct_policy}, one of the options considered for our time interval $\tilde{T}_i$ is $\lceil T^{\max}_i \rceil^{ (\tilde{R}_1) }$. However, by definition,
\[ T^{\max}_i ~~=~~ \max \left\{ \frac{ 1 }{ \eps } \cdot \tilde{T}_{\min}, \sqrt{ \frac{ K_i }{ H_i }  } \right\} ~~=~~ \max \left\{ \frac{ 1 }{ \eps } \cdot \tilde{T}_{\min}, T_i^{ \eoq } \right\} ~~=~~ T_i^{ \eoq } \ , \]
where the last equality holds since $\frac{ 1 }{ \eps } \cdot \tilde{T}_{\min} \leq \frac{ 1 }{ \eps } \cdot T_{\min}^* < T_i^{ \eoq }$, by equation~\eqref{eqn:rel_tildeTmin_Tmin} and the case hypothesis of this regime. Since $\tilde{T}_i$ is picked as the option that minimizes the  EOQ cost $C_i(\cdot)$ of this commodity, 
\begin{eqnarray}
C_i( \tilde{T}_i ) & \leq & C_i( \lceil T^{\max}_i \rceil^{ (\tilde{R}_1) } ) \nonumber \\
& = & C_i( \lceil T^{\eoq}_i \rceil^{ (\tilde{R}_1) } ) \nonumber \\
& \leq & (1 + 4\eps) \cdot C_i( T^{\eoq}_i ) \label{eqn:lem_UB_marginal_hatT_5} \\
& \leq & (1 + 4\eps) \cdot C_i( T^*_i )  \ , \nonumber
\end{eqnarray}
where the last inequality holds since $T^{\eoq}_i$ minimizes $C_i(\cdot)$, as shown in Claim~\ref{clm:EOQ_properties}. To better understand inequality~\eqref{eqn:lem_UB_marginal_hatT_5}, we observe that $\lceil T^{\eoq}_i \rceil^{ (\tilde{R}_1) } \in [T^{\eoq}_i, (1 + 4\eps) \cdot T^{\eoq}_i]$. Indeed, $\lceil T^{\eoq}_i \rceil^{ (\tilde{R}_1) } \geq T^{\eoq}_i$, simply due to rounding up. In the opposite direction, we have 
\begin{eqnarray}
\lceil T^{\eoq}_i \rceil^{ (\tilde{R}_1) } & \leq & T^{\eoq}_i + \tilde{R}_1 \nonumber \\
& \leq & T^{\eoq}_i + (1 + \eps) \cdot R^*_1 \label{eqn:lem_UB_marginal_hatT_6} \\
& \leq & T^{\eoq}_i + (1 + \eps)^2 \cdot T^*_{\min} \label{eqn:lem_UB_marginal_hatT_7} \\
& \leq & (1 + 4\eps) \cdot T^{\eoq}_i \ . \label{eqn:lem_UB_marginal_hatT_8}
\end{eqnarray}
Here, inequality~\eqref{eqn:lem_UB_marginal_hatT_6} holds since $\tilde{R}_1 \in (1 \pm \eps) \cdot R_1^*$, as argued in our proof for the low regime. Inequality~\eqref{eqn:lem_UB_marginal_hatT_7} follows by recalling that $R^*_1 \in S_1^* = [T^*_{\min}, (1 + \eps) \cdot T^*_{\min})$. Finally, inequality~\eqref{eqn:lem_UB_marginal_hatT_8} is obtained by noting that $T_{\min}^* < \eps T_i^{ \eoq }$, due to the case hypothesis of this regime.

%%%%%%%%%%%%%%%%%%%%%%%%%%%%%%%%%
\section{Additional Proofs from Sections~\ref{sec:RC_general}-\ref{sec:RC_constant}} 

% \subsection{Proof of Claim~\ref{clm:bound_round_eU}} \label{app:proof_clm_bound_round_eU}

% \[ \exsub{ U }{ e^U } ~~=~~ \int_0^{ \ln 2 } \frac{ e^u }{ \ln 2 } \mathrm{d}u ~~=~~ \left. \frac{ e^u }{ \ln 2 } \right]_0^{ \ln 2 } ~~=~~ \frac{ 1 }{ \ln 2 } \ . \]
% \[ \exsub{ U }{ e^{-U} } ~~=~~ \int_0^{ \ln 2 } \frac{ e^{-u} }{ \ln 2 } \mathrm{d}u ~~=~~ \left. -\frac{ e^{-u} }{ \ln 2 } \right]_0^{ \ln 2 } ~~=~~ \frac{ 1 }{ 2\ln 2 } \ . \]

\subsection{Proof of Claim~\ref{clm:bound_round_1_3by2}} \label{app:proof_clm_bound_round_1_3by2}

Let us write $T_i^* = \theta T_{\min}^*$, for some $\theta \in [1,\frac{3}{2}]$. With this notation, it is easy to verify that 
\[ \tilde{T}^B_i ~~=~~ 
\begin{cases}
2 \Delta_U, & \text{if } U \in [0,\ln (2/\theta)] \\
3 \Delta_U, & \text{if } U \in (\ln (2/\theta),2] 
\end{cases} \]
Therefore, recalling that $\Delta_U = \frac{ T_{\min}^* }{ e^U } = \frac{ T_i^* }{ \theta e^U }$, we have
\begin{eqnarray*}
    \exsub{ U }{ \frac{ \tilde{T}^B_i }{ T_i^*} } & = & \int_0^{ \ln (2/\theta) } \frac{ 2 }{ \theta e^u \ln 2 } \mathrm{d}u + \int_{ \ln (2/\theta) }^{ \ln 2 } \frac{ 3 }{ \theta e^u\ln 2 } \mathrm{d}u \\
    & = & \left. -\frac{ 2 }{ \theta\ln 2} \cdot e^{-u} \right]_0^{ \ln(2/\theta) } \left. - \frac{ 3 }{ \theta\ln 2} \cdot  e^{-u} \right]_{ \ln(2/\theta) }^{ \ln 2 } \\
    & = & \frac{ 1 }{ \theta\ln 2} \cdot \left( 2 \cdot \left( 1 - \frac{ \theta }{ 2 } \right) + 3 \cdot \left( \frac{ \theta }{ 2 } - \frac{ 1 }{ 2 } \right) \right) \\
    & = & \frac{ 1 }{ 2\ln 2} \cdot \left( 1 +  \frac{ 1 }{ \theta } \right) \\
    & = & \frac{ 1 }{ 2 \ln 2} \cdot \left( 1 +  \frac{ T_{\min}^* }{ T_i^* } \right)
        \ . 
\end{eqnarray*} 

\subsection{Proof of Claim~\ref{clm:bound_round_3by2_2}} \label{app:proof_clm_bound_round_3by2_2}

Following the logic of Appendix~\ref{app:proof_clm_bound_round_1_3by2}, let us again write $T_i^* = \theta T_{\min}^*$, for some $\theta \in (\frac{3}{2},2]$. In this case, 
\[ \tilde{T}^B_i ~~=~~ 
\begin{cases}
2 \Delta_U, & \text{if } U \in [0,\ln (2/\theta)] \\
3 \Delta_U, & \text{if } U \in (\ln (2/\theta),\ln (3/\theta)] \\
4 \Delta_U, & \text{if } U \in (\ln(3/\theta),2] 
\end{cases} \]
Therefore,
\begin{eqnarray*}
    \exsub{ U }{ \frac{ \tilde{T}^B_i }{ T_i^*} } & = & \int_0^{ \ln (2/\theta) } \frac{ 2 }{ \theta e^u \ln 2 } \mathrm{d}u + \int_{ \ln (2/\theta) }^{ \ln (3/\theta) } \frac{ 3 }{ \theta e^u \ln 2 } \mathrm{d}u + \int_{ \ln (3/\theta) }^{ \ln 2 } \frac{ 4 }{ \theta e^u\ln 2 } \mathrm{d}u \\
    & = & \left. -\frac{ 2 }{ \theta\ln 2} \cdot  e^{-u} \right]_0^{ \ln(2/\theta) } - \left. \frac{ 3 }{ \theta\ln 2} \cdot  e^{-u} \right]_{ \ln(2/\theta) }^{ \ln (3/\theta) } \left. - \frac{ 4 }{ \theta\ln 2} \cdot  e^{-u} \right]_{ \ln(3/\theta) }^{ \ln 2 } \\
    & = & \frac{ 1 }{ \theta\ln 2} \cdot \left( 2 \cdot \left( 1 - \frac{ \theta }{ 2 } \right) + 3 \cdot \left( \frac{ \theta }{ 2 } - \frac{ \theta }{ 3 } \right) + 4 \cdot \left( \frac{ \theta }{ 3 } - \frac{ 1 }{ 2 } \right) \right) \\
    & = & \frac{ 5 }{ 6\ln 2} \ . 
\end{eqnarray*} 

\subsection{Proof of Lemma~\ref{lem:UB_prob_large_cost}} \label{app:proof_lem_UB_prob_large_cost}

\paragraph{Cost decomposition.} Clearly, for every commodity $i \in [n]$, there is at most one time interval $t \in {\cal T}_i$ for which $(i,t)$ is an expensive pair chosen by $x^*$, namely, $(i,t) \in {\cal E}^*$; we denote the latter by $t_i$. In addition, let ${\cal N}$ be the set of commodities with such an interval, and let $\bar{\cal N} = [n] \setminus {\cal N}$. Using this notation, the random EOQ cost of our policy is 
\begin{eqnarray*}
\sum_{i \in [n]} \left( \frac{ K_i }{ \tilde{T}_i } + H_i \tilde{T}_i \right) & = & \sum_{ i \in {\cal N} }  \left( \frac{ K_i }{ t_i } + H_i t_i \right) + \sum_{ i \in \bar{\cal N} } \left( \frac{ K_i }{ \tilde{T}_i } + H_i \tilde{T}_i \right) \\
& = & \sum_{ i \in {\cal N} } \sum_{t \in {\cal T}_i} \left( \frac{ K_i }{ t } + H_i t \right) \tilde{x}_{it} + \sum_{ i \in \bar{\cal N} }  \left( \frac{ K_i }{ \tilde{T}_i } + H_i \tilde{T}_i \right) \\
& = & \opt\eqref{eqn:linear_relax_RCJRP} -  \sum_{ i \in \bar{\cal N} } \sum_{t \in {\cal T}_i} \left( \frac{ K_i }{ t } + H_i t \right) \tilde{x}_{it} + \sum_{ i \in \bar{\cal N} }  \left( \frac{ K_i }{ \tilde{T}_i } + H_i \tilde{T}_i \right) \ ,
\end{eqnarray*}
where the first two equalities hold since~\eqref{eqn:linear_relax_RCJRP} includes the constraint ${x}_{it} = 1$ for every $(i,t) \in {\cal E}^*$. Therefore, letting $Z_i = \frac{ K_i }{ \tilde{T}_i } + H_i \tilde{T}_i$, we will conclude the desired claim, $\prpar{ \sum_{i \in [n]} ( \frac{ K_i }{ \tilde{T}_i } + H_i \tilde{T}_i ) \leq (1 + \eps) \cdot \opt\eqref{eqn:linear_relax_RCJRP} } \geq \frac{ 2 }{ 3 }$, by showing that
\begin{equation} \label{eqn:lem_UB_prob_large_cost_1}
\pr{ \sum_{ i \in \bar{\cal N} } Z_i \geq \sum_{ i \in \bar{\cal N} } \sum_{t \in {\cal T}_i} \left( \frac{ K_i }{ t } + H_i t \right) \tilde{x}_{it} + \eps \opt\eqref{eqn:linear_relax_RCJRP} }  ~~\leq~~ \frac{ 1 }{ 3 } \ .    
\end{equation}

\paragraph{Proving inequality~\eqref{eqn:lem_UB_prob_large_cost_1}.} To this end, the important observation is that $\{ Z_i \}_{ i \in \bar{\cal N} }$ are mutually independent, with an expected value of 
\[ \ex{ Z_i } ~~=~~ \ex{ \frac{ K_i }{ \tilde{T}_i } + H_i \tilde{T}_i } ~~=~~ \sum_{t \in {\cal T}_i} \left( \frac{ K_i }{ t } + H_i t \right) \tilde{x}_{it} ~~\leq~~ \eps^4 \opt\eqref{eqn:linear_relax_RCJRP} \ . \]
To better understand the last inequality, note that for every commodity $i \in \bar{\cal N}$, we could have $\tilde{x}_{it} > 0$ only for pairs $(i,t) \notin {\cal E}$, in which case $\frac{ K_i }{ t } + H_i t \leq \eps^4 \opt\eqref{eqn:linear_relax_RCJRP}$. In fact, this observation implies that each such $Z_i$ is upper-bounded by $\eps^4 \opt\eqref{eqn:linear_relax_RCJRP}$ almost surely. Consequently,
\begin{eqnarray*}
&& \pr{ \sum_{ i \in \bar{\cal N} } Z_i \geq \sum_{ i \in \bar{\cal N} } \sum_{t \in {\cal T}_i} \left( \frac{ K_i }{ t } + H_i t \right) \tilde{x}_{it} + \eps \opt\eqref{eqn:linear_relax_RCJRP} } \\
&& \qquad\qquad\qquad\qquad\qquad =~~ \pr{ \sum_{ i \in \bar{\cal N} } Z_i \geq \ex{ \sum_{ i \in \bar{\cal N} } Z_i } + \eps \opt\eqref{eqn:linear_relax_RCJRP} } \ ,    
\end{eqnarray*}
and we proceed by considering two cases, depending on the relation between $\expar{ \sum_{ i \in \bar{\cal N} } Z_i }$ and $\opt\eqref{eqn:linear_relax_RCJRP}$.
\begin{itemize}
    \item {\em When $\expar{ \sum_{ i \in \bar{\cal N} } Z_i } \leq \frac{ \eps }{ 3 } \cdot \opt\eqref{eqn:linear_relax_RCJRP}$:} In this case,
    \begin{eqnarray*}
     \pr{ \sum_{ i \in \bar{\cal N} } Z_i \geq \ex{ \sum_{ i \in \bar{\cal N} } Z_i } + \eps \opt\eqref{eqn:linear_relax_RCJRP} } & \leq & \pr{ \sum_{ i \in \bar{\cal N} } Z_i \geq  \eps \opt\eqref{eqn:linear_relax_RCJRP} } \\
     & \leq & \frac{ \expar{ \sum_{ i \in \bar{\cal N} } Z_i } }{ \eps \opt\eqref{eqn:linear_relax_RCJRP} } \\
     & \leq & \frac{ 1 }{ 3 } \ ,
    \end{eqnarray*}
    where the second inequality is obtained by employing Markov's inequality, and the third inequality follows from our case hypothesis.

    \item {\em When $\expar{ \sum_{ i \in \bar{\cal N} } Z_i } > \frac{ \eps }{ 3 } \cdot \opt\eqref{eqn:linear_relax_RCJRP}$:} In this case, since $\expar{ \sum_{ i \in \bar{\cal N} } Z_i } \leq \opt\eqref{eqn:linear_relax_RCJRP}$, we observe that
    \begin{eqnarray}
    && \pr{ \sum_{ i \in \bar{\cal N} } Z_i \geq \ex{ \sum_{ i \in \bar{\cal N} } Z_i } + \eps \opt\eqref{eqn:linear_relax_RCJRP} } \nonumber \\
    && \qquad \leq~~ \pr{ \sum_{ i \in \bar{\cal N} } Z_i \geq  (1 + \eps) \cdot \ex{ \sum_{ i \in \bar{\cal N} } Z_i } } \nonumber \\
    && \qquad =~~ \pr{ \sum_{ i \in \bar{\cal N} } \frac{ Z_i }{ \eps^4 \opt\eqref{eqn:linear_relax_RCJRP} } \geq  (1 + \eps) \cdot \ex{ \sum_{ i \in \bar{\cal N} } \frac{ Z_i }{ \eps^4 \opt\eqref{eqn:linear_relax_RCJRP} } } } \nonumber \\
     && \qquad \leq~~ \exp \left\{ -\frac{ \eps^2 }{3} \cdot \ex{ \frac{ 1 }{ \eps^4 \opt\eqref{eqn:linear_relax_RCJRP}} \cdot \sum_{ i \in \bar{\cal N} } Z_i }  \right\} \label{eqn:lem_UB_prob_large_cost_2}  \\
     && \qquad \leq~~ \exp \left\{ -\frac{ 1 }{9\eps} \right\} \label{eqn:lem_UB_prob_large_cost_3} \\
     && \qquad \leq~~ \frac{ 1 }{ 3 } \ . \label{eqn:lem_UB_prob_large_cost_4}
    \end{eqnarray}
    Here, to derive inequality~\eqref{eqn:lem_UB_prob_large_cost_2}, we utilize the Chernoff-Hoeffding bound stated by \citet[Thm.~1.1]{DubhashiP09}, noting that $\{ \frac{ Z_i }{ \eps^4 \opt\eqref{eqn:linear_relax_RCJRP} } \}_{ i \in \bar{\cal N} }$ are independent and $[0,1]$-bounded  random variables. Inequality~\eqref{eqn:lem_UB_prob_large_cost_3} follows from the case hypothesis. Finally, inequality~\eqref{eqn:lem_UB_prob_large_cost_4} holds since $e^{ -\frac{ 1 }{9x} } \leq \frac{ 1 }{ 3 }$ for all $x \in (0,\frac{ 1 }{ 10 }]$.    
\end{itemize}

\subsection{Proof of Lemma~\ref{lem:UB_prob_large_resource}} \label{app:proof_lem_UB_prob_large_resource}

\paragraph{Constraint decomposition.} Let us focus on a single constraint $d \in [D]$. Clearly, for every commodity $i \in [n]$, there is at most one time interval $t \in {\cal T}_i$ for which $(i,t)$ is a $d$-heavy pair chosen by $x^*$, namely, $(i,t) \in {\cal H}^*_d$; we denote the latter by $t_i$. In addition, let ${\cal N}$ be the set of commodities with such an interval, and let $\bar{\cal N} = [n] \setminus {\cal N}$. Using this notation, the random resource consumption of our policy with respect to the constraint in question is
\begin{eqnarray*}
\sum_{i \in [n]} \frac{ \alpha_{id} }{ \tilde{T}_i } & = & \sum_{ i \in {\cal N} }  \frac{ \alpha_{id} }{ t_i } + \sum_{ i \in \bar{\cal N} } \frac{ \alpha_{id} }{ \tilde{T}_i } \\
& = & \sum_{ i \in {\cal N} } \alpha_{id} \cdot \sum_{t \in {\cal T}_i} \frac{ \tilde{x}_{it} }{ t } + \sum_{ i \in \bar{\cal N} } \frac{ \alpha_{id} }{ \tilde{T}_i } \\
& \leq & (1 + \eps) \cdot \beta_d - \sum_{ i \in \bar{\cal N} } \alpha_{id} \cdot \sum_{t \in {\cal T}_i} \frac{ \tilde{x}_{it} }{ t } + \sum_{ i \in \bar{\cal N} } \frac{ \alpha_{id} }{ \tilde{T}_i } \ ,
\end{eqnarray*}
with both equalities above holding since~\eqref{eqn:linear_relax_RCJRP} includes the constraint ${x}_{it} = 1$ for every $d \in [D]$ and $(i,t) \in {\cal H}^*_d$. Therefore, letting $Z_i = \frac{ \alpha_{id} }{ \tilde{T}_i }$, we will conclude the desired claim, $\prpar{ \sum_{i \in [n]} \frac{ \alpha_{id} }{ \tilde{T}_i } \geq (1+3\eps) \cdot \beta_d} \leq \frac{ 1 }{ 3D }$, by showing that
\begin{equation} \label{eqn:lem_UB_prob_large_resource_1}
\pr{ \sum_{ i \in \bar{\cal N} } Z_i \geq \sum_{ i \in \bar{\cal N} } \alpha_{id} \cdot \sum_{t \in {\cal T}_i} \frac{ \tilde{x}_{it} }{ t } + 2\eps \beta_d }  ~~\leq~~ \frac{ 1 }{ 3D } \ . 
\end{equation}

\paragraph{Proving inequality~\eqref{eqn:lem_UB_prob_large_resource_1}.} To this end, the important observation is that $\{ Z_i \}_{ i \in \bar{\cal N} }$ are mutually independent, with 
\[ \ex{ Z_i } ~~=~~ \ex{  \frac{ \alpha_{id} }{ \tilde{T}_i } } ~~=~~ \sum_{t \in {\cal T}_i} \frac{ \alpha_{id} }{ t } \cdot \tilde{x}_{it} ~~\leq~~ \delta \beta_d \ . \]
To better understand the last inequality, note that for every commodity $i \in \bar{\cal N}$, we could have $\tilde{x}_{it} > 0$ only for pairs $(i,t) \notin {\cal H}_d$, in which case $\frac{ \alpha_{id} }{ t } \leq \delta \beta_d$. In fact, this observation implies that each such $Z_i$ is upper-bounded by $\delta \beta_d$ almost surely. Consequently,
\[ \pr{ \sum_{ i \in \bar{\cal N} } Z_i \geq \sum_{ i \in \bar{\cal N} } \alpha_{id} \cdot \sum_{t \in {\cal T}_i} \frac{ \tilde{x}_{it} }{ t } + 2\eps \beta_d } ~~=~~ \pr{ \sum_{ i \in \bar{\cal N} } Z_i \geq \ex{ \sum_{ i \in \bar{\cal N} } Z_i } + 2\eps \beta_d } \ , \]
and we proceed by considering two cases, depending on the relation between $\expar{ \sum_{ i \in \bar{\cal N} } Z_i }$ and $\beta_d$:
\begin{itemize}
    \item {\em When $\expar{ \sum_{ i \in \bar{\cal N} } Z_i } \leq \frac{ \eps }{ 3D } \cdot \beta_d$:} In this case,
    \begin{eqnarray*}
     \pr{ \sum_{ i \in \bar{\cal N} } Z_i \geq \ex{ \sum_{ i \in \bar{\cal N} } Z_i } + 2\eps \beta_d } & \leq & \pr{ \sum_{ i \in \bar{\cal N} } Z_i \geq  2\eps \beta_d } \\
     & \leq & \frac{ \expar{ \sum_{ i \in \bar{\cal N} } Z_i } }{ 2\eps \beta_d }  \\
     & \leq & \frac{ 1 }{ 6D } \ ,
    \end{eqnarray*}
    where the second inequality is obtained by employing Markov's inequality, and the third inequality follows from our case hypothesis.

    \item {\em When $\expar{ \sum_{ i \in \bar{\cal N} } Z_i } > \frac{ \eps }{ 3D } \cdot \beta_d$:} In this case, since $\expar{ \sum_{ i \in \bar{\cal N} } Z_i } \leq (1 + \eps) \cdot \beta_d$, we observe that
    \begin{eqnarray}
     \pr{ \sum_{ i \in \bar{\cal N} } Z_i \geq \ex{ \sum_{ i \in \bar{\cal N} } Z_i } + 2\eps \beta_d } & \leq & \pr{ \sum_{ i \in \bar{\cal N} } Z_i \geq  (1 + \eps) \cdot \ex{ \sum_{ i \in \bar{\cal N} } Z_i } } \nonumber \\
     & = & \pr{ \sum_{ i \in \bar{\cal N} } \frac{ Z_i }{ \delta \beta_d } \geq  (1 + \eps) \cdot \ex{ \sum_{ i \in \bar{\cal N} } \frac{ Z_i }{ \delta \beta_d } } } \nonumber \\
     & \leq & \exp \left\{ -\frac{ \eps^2 }{3} \cdot \ex{ \frac{ 1 }{ \delta \beta_d} \cdot \sum_{ i \in \bar{\cal N} } Z_i }  \right\} \label{eqn:lem_UB_prob_large_resource_2}  \\
     & \leq & \exp \left\{ -\frac{ D }{9\eps} \right\} \label{eqn:lem_UB_prob_large_resource_3}\\
     & \leq & \frac{ \eps }{ 3D } \ . \label{eqn:lem_UB_prob_large_resource_4} 
    \end{eqnarray}
    Here, to derive inequality~\eqref{eqn:lem_UB_prob_large_resource_2}, we utilize the Chernoff-Hoeffding bound stated by \citet[Thm.~1.1]{DubhashiP09}, noting that $\{ \frac{ Z_i }{ \delta \beta_d } \}_{ i \in \bar{\cal N} }$ are independent and $[0,1]$-bounded  random variables. Recalling that $\delta = \frac{ \eps^4 }{ D^2 }$, inequality~\eqref{eqn:lem_UB_prob_large_resource_3} follows from the case hypothesis. Finally, inequality~\eqref{eqn:lem_UB_prob_large_resource_4} holds since $e^{ -\frac{ 1 }{9x} } \leq \frac{ x }{ 3 }$ for all $x \in (0,\frac{ 1 }{ 50 }]$. 
\end{itemize}